\documentclass[12pt]{article}
\usepackage[T1]{fontenc}
\usepackage[utf8]{inputenc}
\usepackage[]{frontespizio} %\usepackage[swapnames]{frontespizio}
\usepackage[italian,english]{babel}
\usepackage{xcolor}
\usepackage{hyperref}
\hypersetup{colorlinks=true,
					linktocpage=true,
					linkcolor=red,
					citecolor=green,
					filecolor=cyan,
					urlcolor=magenta}
\usepackage{amsmath}
\usepackage{amssymb}
\usepackage{amsthm}
\usepackage{amsfonts}
\usepackage{graphicx}
\usepackage{float}
\usepackage[left=2.5cm,right=2.5cm,top=3cm,bottom=3cm]{geometry}
\usepackage{setspace}
\usepackage{braket}
\usepackage{slashed}
\usepackage{chemformula}
\usepackage{lscape}
\usepackage{cleveref}
\usepackage{caption}
\usepackage{subfig}
\usepackage[rightcaption]{sidecap}
\usepackage{rotating}
\usepackage{multirow}
\usepackage{tikz-feynman}
\usepackage{pgfplots}
\usetikzlibrary{3d}
\usepackage{cite}

\begin{document}

\begin{flushright}
    BARI-TH/753-24
\end{flushright}

\vspace{0.2cm}

\begin{center}
\Large \textbf{A new look at $b \to s$ observables in 331 models}
\end{center}

%\vspace{0.2cm}

\begin{center}
Francesco Loparco$^a$
\end{center}

%\vspace{0.2cm}

\begin{center}
\small
$^a$Istituto Nazionale di Fisica Nucleare, Sezione di Bari, Via Orabona 4,
I-70126 Bari, Italy
\end{center}

\vspace{0.5cm}

%{\em Version of \today}

\begin{abstract}
\noindent
Flavour changing neutral current (FCNC) processes are described by loop diagrams  in the Standard Model (SM), while in 331 models, based on the gauge group $\text{SU}(3)_C \times \text{SU}(3)_L \times \text{U}(1)_X$, they are dominated by tree-level exchanges of a new heavy neutral gauge boson $Z'$.
Exploiting this feature, observables related to FCNC decays of $K$, $B_d$ and $B_s$ mesons can be considered in several  variants of 331 models.
The variants are distinguished by the value of a parameter $\beta$ that plays a key role in this  framework.
Imposing  constraints on the $\Delta F = 2$ observables, we select possible ranges for the mass of the $Z'$ boson in correspondence to the values  $\beta = \pm k / \sqrt{3}$, with $k = 1, 2$.
The results are used to determine the impact of 331 models on  $b \to s$ processes and on the correlations among them, in the light of new experimental data recently released.
\end{abstract}

\thispagestyle{empty}

\newpage

\section*{Introduction}
\label{sec_intro}

Flavour changing neutral current (FCNC) processes occur in the Standard Model (SM) through loop diagrams, hence they are sensitive to the virtual contribution of heavy particles, even those not yet observed.
For this reason they have played a major role in the search for physics beyond the Standard Model (BSM). 
Among such processes, modes induced by the $b \to s$ transition have been widely investigated in theory and by the experimental Collaborations.
Exclusive $B \to K^{(*)} \, \ell^+ \, \ell^-$, with $\ell=e, \mu$, offer the possibility to exploit a number of observables, such as angular distributions and asymmetries, that are sensitive to BSM.
Ratios of observables are also useful, since the dependence on the Cabibbo-Kobayashi-Maskawa (CKM) matrix elements drops out.
The largest uncertainty is related to the hadronic form factors describing the $B \to K^{(*)}$ matrix elements.
Some observables can be identified where such an uncertainty is largely reduced.
Other interesting modes in this category are $B_s \to \mu^+ \, \mu^-$ and $B \to K^{(*)} \, \nu \, \bar{\nu}$.
Tensions with respect to the SM predictions have emerged in experimental data relative to these modes \cite{HFLAV:2022esi}.

Among the various  scenarios proposed  to extend the SM, promising ones are the 331 models based on the gauge group $\text{SU}(3)_C \times \text{SU}(3)_L \times \text{U}(1)_X$, in which FCNC processes are dominated by tree-level exchanges of a new heavy neutral gauge boson $Z'$.
In this paper we consider how these models face the latest experimental results for a few observables relative to the modes $B \to K^{(*)} \, \ell^+ \, \ell^-$ and $B \to K^{(*)} \, \nu \, \bar{\nu}$.

The plan of the paper is the following.
In Section \ref{sec_transition} we review the main features of the 331 models that are relevant for our study, specifying the classification of the model variants.
A new approach to constrain the 331 parameters  independently of the variant, and to bound the $Z^\prime$ mass in the four considered variants is provided in Section \ref{constraints}.
The SM effective Hamiltonians for $b \to s \, \ell^+ \, \ell^-$ and $b \to s \, \nu \, \bar{\nu}$ are presented in Section \ref{FCNC-Heff}, as well as their modification in the 331 case.
The results of our study are discussed in Section \ref{numerics}.
The last section is devoted to the conclusions.

\section{331 models}
\label{sec_transition}

We briefly describe the class of models  that goes under the name of 331 models \cite{Pisano:1992bxx,Frampton:1992wt}, focusing on the features relevant for our discussion. A detailed description  can be found in \cite{Buras:2012dp}. 

331 models are based on the gauge group $\text{SU}(3)_C \times \text{SU}(3)_L \times \text{U}(1)_X$, first spontaneously broken into the SM one $\text{SU}(3)_C \times \text{SU}(2)_L \times \text{U}(1)_Y$, then to $\text{SU}(3)_C \times \text{U}(1)_Q$.\\
The important differences with respect to the SM, consequences of having enlarged the gauge group, are listed below.
\begin{itemize}
\item
Five new gauge bosons are introduced.
\item
Left-handed fermions can transform according to the fundamental or the conjugate representation of $\text{SU}(3)_L$, i.e. as triplets or antitriplets.
The SM fermions occupy two components of such multiplets, the third component is a new heavy fermion.
Right-handed fermions are singlets as in SM.
The requirement of anomaly cancellation together with the asymptotic freedom of QCD constrains the number of generations and the number of  colours to be equal, so that the 331 models are able to explain the existence of just three generations of fermions.
The same requirement imposes that quark generations  transform differently under $\text{SU}(3)_L$.
One possibility is that two of them  transform as triplets, while the remaining one (generally the third one) as an antitriplet.
The different description of the third generation might explain the origin of the large top quark mass.
In this set up, the leptons transform as antitriplets.
Another possibility is obtained reversing the role of triplets and antitriplets. 
\item
The Higgs sector is extended, and consists of three $\text{SU}(3)_L$ triplets and one sextet.
\item
The electric charge operator $\hat{Q}$ is defined as
\begin{align*}
\hat{Q} = \hat{T}_3 + \beta \, \hat{T}_8 + \hat{X} \;, 
\end{align*}
with $\hat{T}_3$ and $\hat{T}_8$ the diagonal $\text{SU}(3)_L$ generators and $\hat{X}$ the $\text{U}(1)_X$ generator. 
\end{itemize}

Among 331 models, a given variant is specified by a parameter $\beta$ together with the choice between the two possibilities for the fermion representations. 
As a consequence, the charges of the new gauge bosons depend on the variant. However, one of them, usually denoted $Z'$, is  neutral regardless of the value of $\beta$.
This new gauge boson mediates tree-level FCNC in the quark sector, while its couplings to leptons are diagonal and universal.\\
To define quark mass eigenstates, two rotation matrices  are introduced as in the SM.
The one that rotates up-type quarks is denoted as $U_L$, for down-type quarks it is $V_L$.
The relation $V_\text{CKM} = U_L^\dagger \, V_L$ holds, $V_\text{CKM}$ being the CKM matrix.
Differently from the SM, where $V_\text{CKM}$ weights the charged current interactions between up- and down-type quarks and where the two matrices $U_L$ and $V_L$ never appear individually, in 331 models only one between $U_L$ or $V_L$ can be expressed in terms of $V_\text{CKM}$ and of the other one, so that the remaining rotation matrix affects the $Z'$ couplings to  quarks.
Choosing $V_L$ as the remaining matrix,  the following parametrization can be adopted:
\begin{align}
\label{V_L}
V_L =
\left(
\begin{array}{ccc}
\tilde{c}_{12} \, \tilde{c}_{13} & \tilde{s}_{12} \, \tilde{c}_{23} \, e^{i \, \delta_3} - \tilde{c}_{12} \, \tilde{s}_{13} \, \tilde{s}_{23} \, e^{i \, (\delta_1 - \delta_2)} & \tilde{c}_{12} \, \tilde{c}_{23} \, \tilde{s}_{13} \, e^{i \, \delta_1} + \tilde{s}_{12} \, \tilde{s}_{23} \, e^{i \, (\delta_2 + \delta_3)} \\
- \tilde{c}_{13} \, \tilde{s}_{12} \, e^{- i \, \delta_3} & \tilde{c}_{12} \, \tilde{c}_{23} + \tilde{s}_{12} \, \tilde{s}_{13} \, \tilde{s}_{23} \, e^{i \, (\delta_1 - \delta_2 - \delta_3)} & - \tilde{s}_{12} \, \tilde{s}_{13} \, \tilde{c}_{23} \, e^{i \, (\delta_1 - \delta_3)} - \tilde{c}_{12} \, \tilde{s}_{23} \, e^{i \delta_2} \\
- \tilde{s}_{13} \, e^{- i \, \delta_1} & - \tilde{c}_{13} \, \tilde{s}_{23} \, e^{- i \, \delta_2} & \tilde{c}_{13} \, \tilde{c}_{23}
\end{array}
\right) \;,
\end{align}
with $\tilde{c}_i = \cos \theta_i$, $\tilde{s}_i = \sin \theta_i$ and phases $\delta_{1, 2, 3}$.

It is worth  noticing that flavour violating $Z'$ couplings to  quarks involve few parameters in \eqref{V_L} depending on the decay we are considering.
Indeed, the $B_d$ system involves $\tilde{s}_{13}$ and $\delta_1$, the $B_s$ system involves $\tilde{s}_{23}$ and $\delta_2$, the kaon system $\tilde{s}_{13}$, $\tilde{s}_{23}$ and $\delta_2 - \delta_1$, providing a remarkable correlation among the three systems \cite{Buras:2012dp,Buras:2013dea,Buras:2014yna,Buras:2015kwd,Buras:2016dxz,Buras:2023ldz}.
Moreover,  the relation
\begin{align}
\label{U_L}
U_L = V_L \, V_\text{CKM}^\dagger
\end{align}
allows to constrain the $Z'$ mediated FCNC transitions of up-type quark using bounds established in down-type quark sector \cite{Colangelo:2021myn,Buras:2021rdg}.
Such a connection between down-type and up-type quark FCNC processes is a peculiar feature of the 331 models.
The 331 Lagrangian density describing the $Z'$ coupling to ordinary fermions, for a generic value $\beta$, is:
\begin{align}
\label{L_int_331}
i \, \mathcal{L}_\text{int}^{Z'} & = i \, \frac{g \, Z^{\prime \mu}}{2 \, \sqrt{3} \, c_W \, \sqrt{1 - (1 + \beta^2) \, s_W^2}} \times \notag \\[1ex]
& \times \Bigg\{ \sum_{\ell = e, \mu, \tau} \, \Big\{ \left[ 1 - (1 + \sqrt{3} \, \beta) \, s_W^2 \right] \,\big( \bar{\nu}_{\ell L} \, \gamma_\mu \, \nu_{\ell L} + \bar{\ell}_L \, \gamma_\mu \, \ell_L \big) - 2 \, \sqrt{3} \, \beta \, s_W^2 \, \bar{\ell}_R \, \gamma_\mu \, \ell_R + \notag \\[1ex]
%& \qquad - \left[ 2 - \left( 2 - \sqrt{3} \, \beta \, (1 - \sqrt{3} \, \beta) \right) \, s_W^2 \right] \, \bar{E}_{\ell L} \, \gamma_\mu \, E_{\ell L} - \sqrt{3} \, (1 - \sqrt{3} \, \beta) \, \beta \, s_W^2 \, \bar{E}_{\ell R} \, \gamma_\mu \, E_{\ell R} \Big\} + \notag \\[1ex]
& \hspace{0.5cm} + \sum_{i, j = 1, 2, 3} \, \bigg\{ \left[ -1 + \left( 1 + \frac{\beta}{\sqrt{3}} \right) \, s_W^2 \right] \, \big( \bar{q}_{uL} \big)_i \, \gamma_\mu \, \big( q_{uL} \big)_j \, \delta_{ij} + 2 \, c_W^2 \, \big( \bar{q}_{uL} \big)_i \, \gamma_\mu \, \big( q_{uL} \big)_j \, u_{3i}^* \, u_{3j} + \notag \\[1ex]
& \hspace{1cm} + \left[ -1 + \left( 1 + \frac{\beta}{\sqrt{3}} \right) \, s_W^2 \right] \, \big( \bar{q}_{dL} \big)_i \, \gamma_\mu \, \big( q_{dL} \big)_j \, \delta_{ij} + 2 \, c_W^2 \, \big( \bar{q}_{dL} \big)_i \, \gamma_\mu \, \big( q_{dL} \big)_j \, v_{3i}^* \, v_{3j} + \notag \\[1ex]
& \hspace{1cm} + \frac{4}{\sqrt{3}} \, \beta \, s_W^2 \, \big( \bar{q}_{uR} \big)_i \, \gamma_\mu \, \big( q_{uR} \big)_j \, \delta_{ij} - \frac{2}{\sqrt{3}} \, \beta \, s_W^2 \, \big( \bar{q}_{dR} \big)_i \, \gamma_\mu \, \big( q_{dR} \big)_j \, \delta_{ij} \Bigg\} \;,
%& \qquad + \sum_{i = 1, 2} \, \bigg\{ \left[ 2 + \frac{- 2 \, \sqrt{3} + \beta \, (1 - 3 \, \sqrt{3} \, \beta)}{\sqrt{3}} \, s_W^2 \right] \, \big( \bar{Q}_{D L} \big)_i \, \gamma_\mu \, \big( Q_{D L} \big)_i + \notag \\[1ex]
%& \qquad \qquad + \frac{\beta \, (1 - 3 \, \sqrt{3} \, \beta)}{\sqrt{3}} \, s_W^2 \, \big( \bar{Q}_{DR} \big)_i \, \gamma_\mu \, \big( Q_{DR} \big)_i  \bigg\} + \notag \\[1ex]
%& \qquad \qquad + \left[ - 2 + \frac{2 \, \sqrt{3} + \beta \, (1 + 3 \, \sqrt{3} \, \beta)}{\sqrt{3}} \, s_W^2 \right] \, \bar{T}_L \, \gamma_\mu \, T_L + \frac{\beta \, (1 + 3 \, \sqrt{3} \, \beta)}{\sqrt{3}} \, s_W^2 \, \bar{T}_R \, \gamma_\mu \, T_R \Bigg\} \;.
\end{align}
with $s_W = \sin \theta_W$ and $c_W = \cos \theta_W$, $q_u$ ($q_d$) denoting an up-type (down-type) quark ($i$, $j$ are generation indices), and $u_{ij}$ and $v_{ij}$ are, respectively, the elements of $U_L$ and $V_L$ matrices.
Following \cite{Buras:2012dp} we write 
the $Z'$ couplings to down-type quarks as:
\begin{align}
\label{eqZpq}
i \, \mathcal{L}_{L}(Z') = i \, \left[ \Delta_L^{sd}(Z') \, ( \bar{s} \, \gamma^\mu \, P_L \, d )+\Delta_L^{bd}(Z') \, ( \bar{b} \, \gamma^\mu \, P_L \, d ) + \Delta_L^{bs}(Z') \, ( \bar{b} \, \gamma^\mu \, P_L \, s ) \right] \, Z'_\mu \;,
\end{align}
the first upper index  denotes the outgoing quark, while the second index the  incoming one, so that:
\begin{align}
\Delta_L^{ji}(Z') = [ \Delta_L^{ij}(Z') ]^* \;.
\end{align}
The $Z'$ couplings to leptons are defined analogously; they are denoted as $\Delta_L^{\nu \bar{\nu}}(Z')$ and $\Delta_{L,R}^{\mu \bar{\mu}}(Z')$.
We list these couplings below, since they play a central role in our discussion:
\begin{subequations}
\label{C_L_331_interi}
\begin{align}
\Delta_L^{sd}(Z') & = \frac{g \, c_W}{\sqrt{3}} \, \sqrt{f(\beta)} \, v_{32}^* \, v_{31} \;, \\
\Delta_L^{bd}(Z') & = \frac{g \, c_W}{\sqrt{3}} \, \sqrt{f(\beta)} \, v_{33}^* \, v_{31} \;, \\
\Delta_L^{bs}(Z') & = \frac{g \, c_W}{\sqrt{3}} \, \sqrt{f(\beta)} \, v_{33}^* \, v_{32} \;,
%\Delta_L^{uc}(Z') & = \frac{g \, c_W}{\sqrt{3}} \, \sqrt{f(\beta)} \, u_{31}^* \, u_{32} \;,
\end{align}
\end{subequations}
%\vspace{-0.4cm}
\begin{subequations}
\label{Delta_for_leptons}
\begin{align}
\Delta_L^{\mu \bar\mu}(Z') & = \Delta_L^{\nu \bar{\nu}}(Z') = \frac{g \, \left[ 1 - ( 1 + \sqrt{3} \, \beta ) \, s_W^2 \right]}{2 \, \sqrt{3} \, c_W \, \sqrt{1 - ( 1 + \beta^2 ) \, s_W^2}} \;, \\
\Delta_R^{\mu \bar\mu}(Z') & =
\left\{
\begin{array}{ll}
\frac{-g \, \beta \, s_W^2}{c_W \, \sqrt{1 - ( 1 + \beta^2 ) \, s_W^2}} \hspace{0.5cm} \text{for} \hspace{0.5cm} \beta \neq \sqrt{3} \\
\frac{g \sqrt{1 - 4 \, s_W^2}}{\sqrt{3} \, c_W} \hspace{1.6cm} \text{for} \hspace{0.5cm} \beta = \sqrt{3}
\end{array}
\right.
\;.
\end{align}
\end{subequations}
The $Z - Z'$ mixing can be considered, with an impact on several flavour observables, however it is negligible in $\Delta F = 2$ transitions considered below.
The $Z - Z'$ mixing angle is written as \cite{Buras:2014yna}.
\begin{equation}
\label{sin_mixing}
\sin \xi = \frac{c_W^2}{3} \, \sqrt{f(\beta)} \, \left( 3 \,  \beta \, \frac{s_W^2}{c_W^2} + \sqrt{3} \, a \right) \, \left[ \frac{M_Z^2}{M_{Z'}^2} \right] \equiv B(\beta, a) \, \left[ \frac{M_Z^2}{M_{Z'}^2} \right] \;,
\end{equation}
where
\begin{equation}
\label{f(beta)}
f(\beta) = \frac{1}{1 - ( 1 + \beta^2 ) \, s_W^2} > 0 \;.
\end{equation}
The parameter $a$ introduced in eq. \eqref{sin_mixing} is defined  as follows:
\begin{equation}
-1 < a = \frac{v_-^2}{v_+^2} < 1 \;,
\end{equation}
with $v_\pm$ related to the VEVs of two Higgs triplets $\rho$ and $\eta$,
\begin{align}
v_+^2 = v_\eta^2 + v_\rho^2 \;, \hspace{1cm} v_-^2 = v_\eta^2 - v_\rho^2 \;.
\end{align}
Adopting a notation similar to the two Higgs doublet models, $a$ can be written in terms of the parameter $\tan \bar{\beta}$ as
\begin{equation}
\label{a_parameter}
a = \frac{1 - \tan^2 \bar{\beta}}{1 + \tan^2 \bar{\beta}} \;, \hspace{1cm} \tan \bar{\beta} = \frac{v_\rho}{v_\eta} \;.
\end{equation}
Finally,  the $\text{U}(1)_X$ gauge coupling $g_X$ and the $\text{SU}(3)_L$ coupling $g$ obey the relation
\begin{align}
\label{ggX}
\frac{g_X^2}{g^2} = \frac{6 \, \sin^2 \theta_W}{1 - ( 1 + \beta^2 ) \, \sin^2 \theta_W} \;. 
\end{align}
Depending on the different values of $\beta$ and $\tan \bar{\beta}$, it is possible to define twenty-four 331 models, called $M_i$ with $i = \{ 1, 2, \dots, 24 \}$. 
Each one of them is analyzed in two scenarios, called $F_1$ and $F_2$, where $F_1$ stands for the case having two generations of quarks belonging to triplets of $\text{SU}(3)_L$ and the third generation to antitriplets; otherwise for $F_2$.
4 out of 24 models have been selected in the present analysis, corresponding to the $F_1$ scenario and $\tan \bar{\beta} = 1$.
They are \cite{Buras:2023ldz}:
\begin{subequations}
\begin{align}
M_1 \hspace{1cm} \longleftrightarrow \hspace{1cm} \beta = - 2 / \sqrt{3} \;, \\
M_3 \hspace{1cm} \longleftrightarrow \hspace{1cm} \beta = - 1 / \sqrt{3} \;, \\
M_5 \hspace{1cm} \longleftrightarrow \hspace{1cm} \beta = + 1 / \sqrt{3} \;, \\
M_7 \hspace{1cm} \longleftrightarrow \hspace{1cm} \beta = + 2 / \sqrt{3} \;.
\end{align}
\end{subequations}

\section{Constraining the 331 parameters independently of the variant}
\label{constraints}

We have stressed that the parameter $\beta$ defines the specific 331 model.
Indeed, it cannot assume arbitrary values.
The following observations constrain its possible values:
\begin{itemize}
\item
requiring that the four new gauge bosons that are introduced together with $Z'$ have integer electric charges constrains the values of $\beta$ to be a multiple of $1 / \sqrt{3}$ or $\sqrt{3}$;
\item
eq. \eqref{ggX} provides the bound
\begin{align}
| \beta | \leq \frac{1}{\tan \theta_W ( M_{Z'} )} \;,
\end{align}
%\noindent
which corresponds to $| \beta | < 1.737$ when $\sin \theta_W ( M_{Z'} \simeq 1 \, \text{TeV} ) = 0.249$.
\end{itemize}
The two observations restrict the allowed values of $\beta$ to $\pm \frac{1}{\sqrt{3}}$, $\pm\frac{2}{\sqrt{3}}$ and $\pm \sqrt{3}$.
However, as observed in \cite{Buras:2013dea}, 331 models with $\beta=\pm \sqrt{3}$ are characterized by a Landau singularity when $\sin^2 \theta_W \simeq 0.25$. This value is reached through the renormalization group evolution at the scale $M_{Z'}\simeq 4$ TeV.
On the other hand, for $| \beta | < \sqrt{3} - 0.2$ the theory is free of Landau singularity up to the GUT scales.

Existing analyses of flavour observables in 331 models usually adopt the strategy of finding allowed values for the parameters $\tilde{s}_{13}$, $\delta_1$, $\tilde{s}_{23}$ and $\delta_2$ in correspondence of the selected values of $M_{Z'}$ and for several values of $\beta$.
The allowed regions are selected imposing that the experimental ranges for $\Delta F=2$ observables are reproduced.
Such observables are the mass differences between neutral mesons, $\Delta M_d$, $\Delta M_s$ and $\Delta M_K$; the CP asymmetries in neutral $B$ meson decays, i.e. $S_{J/\psi K_S}$ in the case of $B_d$ and $S_{J/\psi \phi}$ for $B_s$; the CP violating parameter $\epsilon_K$ for kaon system.
In this study we wish to understand whether flavour observables can give us some information \textit{a priori} on $M_{Z'}$ and $\beta$.
Therefore, we adopt a strategy, described below, in which $M_{Z'}$ and  $\beta$ are left free until the end of the procedure.

The SM contributions to the off-diagonal elements $M_{12}^i$ for neutral $K$ and $B_q$ meson mass matrices read
\begin{subequations}
\begin{align}
\big( M_{12}^K \big)_\text{SM}^* & = \frac{G_F^2}{12 \, \pi^2} \, F_K^2 \, \hat{B}_K \, M_K \, M_W^2 \, \Big[ ( \lambda_c^{(K)} )^2 \, \eta_1 \, S_0(x_c) + ( \lambda_t^{(K)} )^2 \, \eta_2 \, S_0(x_t) + \notag \\[0.5ex]
& \hspace{5cm} + 2 \, \lambda_c^{(K)} \, \lambda_t^{(K)} \, \eta_3 \, S_0(x_c,x_t) \Big] \;, \\
\big( M_{12}^q \big)_\text{SM}^* & = \frac{G_F^2}{12 \, \pi^2} \, F_{B_q}^2 \, \hat{B}_{B_q} \, M_{B_q} \, M_W^2 \, \Big[ ( \lambda_t^{(q)} )^2 \, \eta_B \, S_0(x_t) \Big] \;,
\end{align}
\end{subequations}
where $G_F$ is the Fermi constant, $x_i = m_i^2 / M_W^2$ and
\begin{align}
\lambda_i^{(K)} = V_{is}^* \, V_{id} \hspace{0.5cm} \text{and} \hspace{0.5cm} \lambda_i^{(q)} = V_{tb}^* \, V_{tq} \;,
\end{align}
with $V_{ij}$ the CKM matrix element.
$S_0(x_i)$ and $S_0(x_c, x_t)$ are one-loop box functions that can be found e.g. in \cite{Blanke:2006sb}, while the factors $\eta_i$ are QCD corrections evaluated at the NLO in \cite{Herrlich:1993yv,Herrlich:1995hh,Herrlich:1996vf,Buras:1990fn,Urban:1997gw} and, for $\eta_1$ and $\eta_3$, at NNLO in \cite{Brod:2011ty,Buras:2012fs}.
$\eta_B$ can be found in \cite{Buras:1990fn,Urban:1997gw}.
$\hat{B}_K$ and $\hat{B}_{B_q}$ are the $K$ and $B_q$ meson bag parameters, that are non-perturbative quantities, while $F_K$ and $F_{B_q}$ are $K$ and $B_q$ decay constants, respectively.
In 331 models the flavour independent $S_0(x_t)$ functions must be replaced  by  the functions $S_i$ ($i = K, B_d, B_s$):
\begin{align}
S_i = S_0(x_t) + \Delta S_i(Z') + \Delta S_i(\text{Box}) \equiv | S_i | \, e^{i \, \theta_S^i}
\end{align}
which have  important properties.
In fact, they depend on  flavour, i.e. they are different for the three considered systems,  and carry a new complex phase.
Therefore, we find the results
\begin{subequations}
\begin{align}
\big( M_{12}^K \big)_\text{NP}^* & = \frac{G_F^2}{12 \, \pi^2} \, F_K^2 \, \hat{B}_K \, M_K \, M_W^2 \, \big[ ( \lambda_t^{(K)} )^2 \, \eta_2 \big] \, \bigg[ \frac{\Delta_L^{sd}(Z')}{\lambda_t^{(K)}} \bigg]^2 \, \frac{4 \, \tilde{r}}{M_{Z'}^2 \, g_\text{SM}^2} \;, \\
\big( M_{12}^q \big)_\text{NP}^* & = \frac{G_F^2}{12 \, \pi^2} \, F_{B_q}^2 \, \hat{B}_{B_q} \, M_{B_q} \, M_W^2 \, \big[ ( \lambda_t^{(q)} )^2 \, \eta_B \big] \, \bigg[ \frac{\Delta_L^{bq}(Z')}{\lambda_t^{(q)}} \bigg]^2 \, \frac{4 \, \tilde{r}}{M_{Z'}^2 \, g_\text{SM}^2} \;.
\end{align}
\end{subequations}
where
\begin{align}
\label{gSM2}
g_\text{SM}^2 = 4 \, \frac{G_F}{\sqrt{2}} \, \frac{\alpha}{2 \, \pi \, \sin^2 \theta_W(Z)}
\end{align}
while $\tilde{r}$ can be found in \cite{Buras:2012dp}.
We obtain
\begin{subequations}
\label{starting_point}
\begin{align}
\Delta M_K & = 2 \, \big[ \text{Re} \big( M_{12}^K \big)_\text{SM} + \text{Re} \big( M_{12}^K \big)_\text{NP} \big] \;, \\
\epsilon_K & = \frac{k_\epsilon \, e^{i \, \varphi_\epsilon}}{\sqrt{2} \, \big( \Delta M_K \big)_\text{exp}} \, \big[ \text{Im} \big( M_{12}^K \big)_\text{SM} + \text{Im} \big( M_{12}^K \big)_\text{NP} \big] \;, \\
\Delta M_q & = 2 \, \big| \big( M_{12}^q \big)_\text{SM} + \big( M_{12}^q \big)_\text{NP} \big| \;, \\
S_{J/\psi K_s} & = \sin ( 2 \, \beta + 2 \, \varphi_{B_d} ) \;, \\
S_{J/\psi \phi} & = \sin ( 2 \, | \beta_s | - 2 \, \varphi_{B_s} ) \;,
\end{align}
\end{subequations}
where $ \varphi_{B_q}$ are the phases of $M_{12}^q = (M_{12})_\text{SM}+(M_{12})_\text{NP}$.
Moreover we have  $\varphi_\epsilon = ( 43.51 \pm 0.05 )^\circ$ and $\kappa_\epsilon = 0.94 \pm 0.02$ obtained in \cite{Buras:2008nn,Buras:2010pza}.

The quantities in \eqref{starting_point} can be written in the compact form
\begin{subequations}
\label{starting_point_refined}
\begin{align}
\Delta M_K & = \kappa_1 \, \big( \kappa_2 + \kappa_3 \, n \, \text{Re} \big[ \tilde{K} \big] \big) \;, \\
| \epsilon_K | & = \kappa_4 \, \big| \kappa_5 + \kappa_3 \, n \, \text{Im} \big[ \tilde{K} \big] \big| \;, \\
\Delta M_d & = \Delta_1 \, \sqrt{\big( \Delta_2 + \Delta_3 \, n \, \text{Re} \big[ \tilde{D} \big] \big)^2 + \big( \Delta_4 + \Delta_3 \, n \, \text{Im} \big[ \tilde{D} \big] \big)^2} \;, \\
S_{J/\psi K_s} & = \frac{\Delta_4 + \Delta_3 \, n \, \text{Im} \big[ \tilde{D} \big]}{\sqrt{\big( \Delta_2 + \Delta_3 \, n \, \text{Re} \big[ \tilde{D} \big] \big)^2 + \big( \Delta_4 + \Delta_3 \, n \, \text{Im} \big[ \tilde{D} \big] \big)^2}} \;, \\
\Delta M_s & = \Sigma_1 \, \sqrt{\big( \Sigma_2 + \Sigma_3 \, n \, \text{Re} \big[ \tilde{S} \big] \big)^2 + \big( - \Sigma_4 + \Sigma_3 \, n \, \text{Im} \big[ \tilde{S} \big] \big)^2} \;, \\
S_{J/\psi \phi} & = - \frac{- \Sigma_4 + \Sigma_3 \, n \, \text{Im} \big[ \tilde{S} \big]}{\sqrt{\big( \Sigma_2 + \Sigma_3 \, n \, \text{Re} \big[ \tilde{S} \big] \big)^2 + \big( - \Sigma_4 + \Sigma_3 \, n \, \text{Im} \big[ \tilde{S} \big] \big)^2}} \;,
\end{align}
\end{subequations}
where $\kappa_i$, $\Delta_i$ and $\Sigma_i$ are all positive quantities, $\Delta_1 = \Sigma_1$, and
\begin{subequations}
\label{variabili_tildate}
\begin{align}
\tilde{K} & = e^{2 \, i \, ( \delta_1 - \delta_2 )} \, \tilde{s}_{13}^2 \, ( 1 - \tilde{s}_{13}^2 ) \, \tilde{s}_{23}^2 = e^{2 \, i \, ( \delta_1 - \delta_2 )} \, \tilde{y}_k \;, \\
\tilde{D} & = e^{2 \, i \, \delta_1} \, \tilde{s}_{13}^2 \, ( 1 - \tilde{s}_{13}^2 ) \, ( 1 - \tilde{s}_{23}^2 ) = e^{2 \, i \, \delta_1} \, \tilde{y}_d \;, \\
\tilde{S} & = e^{2 \, i \, \delta_2} \, ( 1 - \tilde{s}_{13}^2 )^2 \,\tilde{s}_{23}^2 \, ( 1 - \tilde{s}_{23}^2 ) = e^{2 \, i \, \delta_2} \, \tilde{y}_s \;.
\end{align}
\end{subequations}
We have introduced the quantity
\begin{align}
n \equiv n(M_{Z'}^2,\beta^2) = \frac{\tilde{r}}{M_{Z'}^2} \, f(\beta) \hspace{1cm} \text{where} \hspace{1cm} n > 0 \;.
\end{align}
%which satisfies  $n > 0$.

\begin{figure}[!tb]
\begin{center}
\includegraphics[scale=0.6]{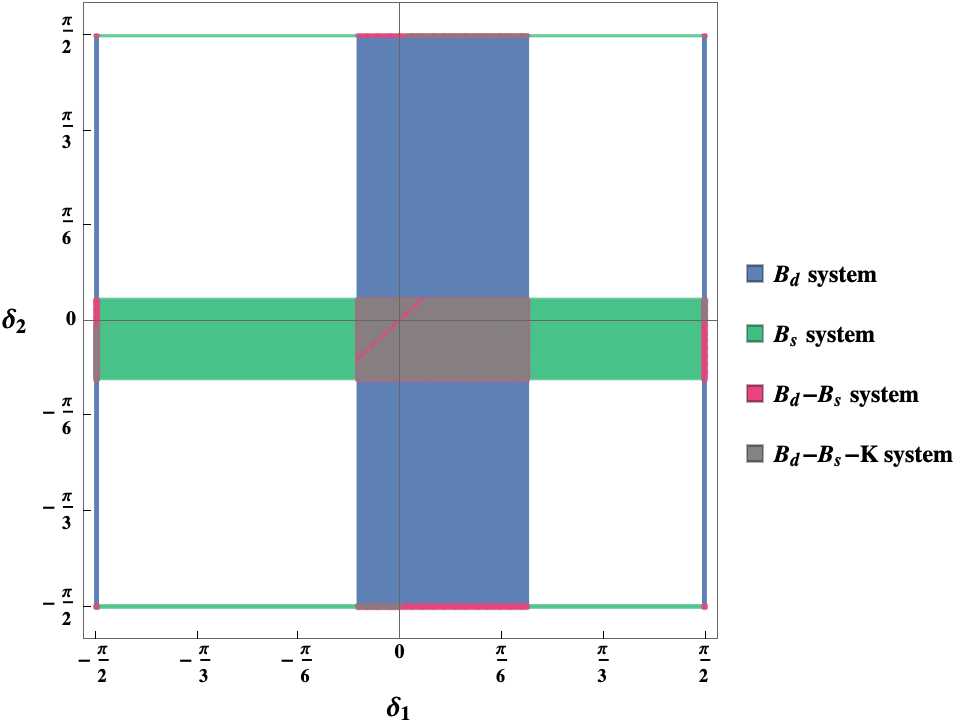}
\end{center}
\vspace{-0.5cm}
\caption{
\small
Allowed ranges for $\delta_1$ and $\delta_2$.
The blue (green) region is obtained using the conditions on the observables relative to the $B_d$ ($B_s$) system.
The red region is the intersection of those regions, while the gray one is obtained taking into account also the constraints from the observables in the $K$ system.
}
\label{delta1_VS_delta2}
\end{figure}

The model parameters are selected imposing that $\Delta M_{B_d}$, $S_{J/\psi K_S}$, $\Delta M_{B_s}$, $S_{J/\psi \phi}$ and $| \epsilon_K |$ lie in their experimental ranges within $2 \, \sigma$.
For $\Delta M_K$ we impose that it lies in the range $[0.75, 1.25] \times ( \Delta M_K )_\text{SM}$, with $( \Delta M_K )_\text{SM} = 4.7 \times 10^{-3} \, \text{GeV}$.
All the input parameters are in Tab. \ref{tab_entries}.
\begin{table}[!tb]
\centering
\small
\begin{tabular}{|l|l|}
\hline
\multicolumn{2}{|c|}{SM parameters}
\\
\hline
$G_F = 1.16637(1) \times 10^{-5} \, \text{GeV}^{-2}$ \hfill \cite{Zyla:2020ssz}
&
$m_c(m_c) = 1.279(13) \, \text{GeV}$ \hfill \cite{Chetyrkin:2017lif}
\\
$M_W = 80.385(15) \, \text{GeV}$ \hfill \cite{Zyla:2020ssz}
&
$m_b(m_b) = 4.163(16) \, \text{GeV}$ \hfill \cite{Chetyrkin:2009fv,Zyla:2020ssz}
\\
$\sin^2 \theta_W = 0.23121(4)$ \hfill \cite{Zyla:2020ssz}
&
$m_t(m_t) = 162.83(67) \, \text{GeV}$ \hfill \cite{Brod:2021hsj}
\\
$\alpha(M_Z) = 1/127.9$ \hfill \cite{Zyla:2020ssz}
&
\\
$\alpha_s^{(5)}(M_Z)= 0.1179(10) $ \hfill \cite{Zyla:2020ssz}
&
\\
\hline
\multicolumn{2}{|c|}{$K$ meson}
\\
\hline
$m_{K^+} = 493.677(13) \, \text{MeV}$ \hfill\cite{Zyla:2020ssz}
&
$\Delta M_K = 0.005292(9) \,\text{ps}^{-1}$ \hfill \cite{Zyla:2020ssz}
\\
$\tau(K^+) = 1.2380(20)\times 10^{-8} \,\text{s}$ \hfill \cite{Zyla:2020ssz}
&
$| \epsilon_K | = 2.228(11) \times 10^{-3}$ \hfill \cite{Zyla:2020ssz}
\\
$m_{K^0} = 497.61(1) \, \text{MeV}$ \hfill \cite{Zyla:2020ssz}
&
$F_K = 155.7(3) \, \text{MeV}$ \hfill \cite{Aoki:2019iem}
\\
$\tau(K_S) = 0.8954(4) \times 10^{-10} \,\text{s}$ \hfill \cite{Zyla:2020ssz}
&
$\hat B_K = 0.7625(97)$ \hfill \cite{Aoki:2019iem}
\\
$\tau(K_L) = 5.116(21) \times 10^{-8} \,\text{s}$ \hfill \cite{Zyla:2020ssz}
&
\\
\hline
\multicolumn{2}{|c|}{$B_d$ meson}
\\
\hline
$m_{B_d} = 5279.58(17) \,\text{MeV}$ \hfill \cite{Zyla:2020ssz}
&
$\Delta M_d = 0.5065(19) \,\text{ps}^{-1}$ \hfill \cite{Zyla:2020ssz}
\\
$\tau(B_d)= 1.519(4) \,\text{ps}$ \hfill \cite{HFLAV:2016hnz}
&
$S_{J/\psi K_s} = 0.699(17)$ \hfill \cite{Zyla:2020ssz}
\\
&
$F_{B_d} = 190.0(1.3) \, \text{MeV}$ \hfill \cite{FlavourLatticeAveragingGroupFLAG:2021npn}
\\
&
$\hat B_{B_d} = 1.222(61)$ \hfill \cite{Dowdall:2019bea}
\\
&
$F_{B_d} \sqrt{\hat B_{B_d}} = 210.6(5.5) \, \text{MeV}$ \hfill \cite{Dowdall:2019bea}
\\
\hline
\multicolumn{2}{|c|}{$B_s$ meson}
\\
\hline
$m_{B_s} = 5366.8(2) \,\text{MeV}$ \hfill \cite{Zyla:2020ssz}
&
$\Delta M_s = 17.749(20) \,\text{ps}^{-1}$ \hfill \cite{Zyla:2020ssz}
\\
$\tau(B_s)= 1.515(4) \,\text{ps}$ \hfill \cite{HFLAV:2016hnz}
&
$S_{J/\psi \phi} = 0.054(20)$ \hfill \cite{Aoki:2019iem}
\\
&
$F_{B_s} = 230.3(1.3) \, \text{MeV}$ \hfill \cite{FlavourLatticeAveragingGroupFLAG:2021npn}
\\
&
$\hat B_{B_s} = 1.232(53)$ \hfill \cite{Dowdall:2019bea}
\\
&
$F_{B_s} \sqrt{\hat B_{B_s}} = 256.1(5.7) \, \text{MeV}$ \hfill \cite{Dowdall:2019bea}
\\
\hline
\multicolumn{2}{|c|}{CKM parameters}
\\
\hline
$| V_{us} | = 0.2253(8)$ \hfill \cite{Zyla:2020ssz}
&
$| V_{cd} | = 0.22517$
\\
$| V_{cb} | = (41.0 \pm 1.4) \times 10^{-3}$ \hfill \cite{Zyla:2020ssz}
&
$| V_{cs} | = 0.97346$
\\
$| V_{ub} | = 3.72 \times 10^{-3}$ \hfill \cite{Zyla:2020ssz}
&
$| V_{td} | = 0.00857$
\\
$\gamma = 68^\circ$ \hfill \cite{Zyla:2020ssz}
&
$| V_{ts} | = 0.04027$
\\
\hline
\end{tabular}
\vspace{0.2cm}
\caption{
\small
Parameters used in the analysis.
}
\label{tab_entries}
\end{table}
For each value of $\beta$ we find the allowed range of $M_{Z'}$, which means that outside such a range, for no values of the 331 parameters $\tilde{s}_{13}$, $\tilde{s}_{23}$, $\delta_1$ and $\delta_2$ the experimental data on the $\Delta F=2$ flavour observables can be reproduced.
In Fig. \ref{delta1_VS_delta2} we show the allowed regions obtained for the two phases $\delta_1$ and $\delta_2$, while the results for the masses of the $Z'$ boson are summarized below:
\begin{subequations}
\begin{align}
& \beta = \pm 1 / \sqrt{3} \hspace{1cm} \longrightarrow \hspace{1cm} M_{Z'} \in [ 1.2, 47] \,\text{TeV} \;, \\
& \beta = \pm 2 / \sqrt{3} \hspace{1cm} \longrightarrow \hspace{1cm} M_{Z'} \in [ 1.5, 59] \,\text{TeV} \;, \\
& \beta = \pm \sqrt{3} \hspace{1.4cm} \longrightarrow \hspace{1cm} M_{Z'} \in [ 14.7, 590] \,\text{TeV} \;.
\end{align}
\end{subequations}
%{
%\color{red}
Even though the model with $\beta = \pm \sqrt{3}$ is excluded due to the Landau singularity, we have computed the allowed $M_{Z'}$ range for completeness.
%We compute the $M_{Z'}$ range for $\beta = \pm \sqrt{3}$ for completeness.
%Notice that it has a different order of magnitude with respect to those obtained for $\beta = \pm k / \sqrt{3}$, with $k = 1, 2$, because of the Landau singularity described above.
%}

\section{FCNC processes: effective Hamiltonian in SM and 331 models}
\label{FCNC-Heff}

\begin{figure}[!tb]
\begin{center}
\begin{tikzpicture}
\begin{feynman}
\vertex (x2);
\vertex [above left=of x2] (x1) {$b$};
\vertex [below left=of x2] (x3) {$s$};
\vertex [right=of x2] (x4);
\vertex [above right=of x4] (x5) {$\ell$};
\vertex [below right=of x4] (x6) {$\bar{\ell}$};
\vertex at ($(x2)!-1.5cm!(x4)!-0.0cm!90:(x4)$) {$\Delta_L^{bs}(Z')$};
\vertex at ($(x2)!+3.0cm!(x4)!-0.0cm!90:(x4)$) {$\Delta_L^{\ell\bar{\ell}}(Z')$};
 
\diagram* {
	(x1) -- [plain, thick, with arrow=0.5,arrow size=0.15em] (x2),
	(x2) -- [plain, thick, with arrow=0.5,arrow size=0.15em] (x3),
	(x4) -- [plain, thick, with arrow=0.5,arrow size=0.15em] (x5),
	(x6) -- [plain, thick, with arrow=0.5,arrow size=0.15em] (x4),
	(x2) -- [boson, thick, edge label={$Z'$}] (x4),
	};
\end{feynman}
\end{tikzpicture}
\end{center}
\vspace{-0.6cm}
\caption{
\small
Feynman diagram for $b \to s \, \ell \, \bar{\ell}$ transition mediated by $Z'$ in the 331 models, with $\ell = \{\mu, \nu\}$}
\label{fig_tree_level}
\end{figure}
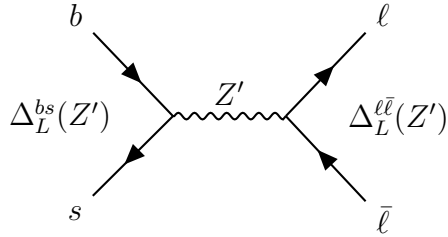

To reconsider the rare FCNC $B$ decays in the framework of 331 models, we recall the SM effective Hamiltonian describing such modes and how it is modified in the 331 case.
Specifically, we are concerned with $b \to s \, \ell^+ \, \ell^-$ and $b \to s \, \nu \, \bar{\nu}$ modes.
The contribution from the tree level $Z'$ exchange is depicted in Fig. \ref{fig_tree_level}.

In SM the effective Hamiltonian governing $b \to s \, \ell^+ \, \ell^-$ reads \cite{Buras:2020xsm}:
\begin{align}
\label{hamil}
\mathcal{H}^\text{SM, eff}_{b \to s \, \ell^+ \, \ell^-} = - 4 \,\frac{G_F }{\sqrt{2}} \, V_{tb} \, V_{ts}^* \, \Big\{C_1 \, O_1 + C_2 \, O_2 + \sum_{i = 3,\dots,6} \, C_i \, O_i + \sum_{i = 7,\dots,10} \, C_i \, O_i \Big\} + \text{h.c.} \;.
\end{align}
Doubly Cabibbo suppressed terms proportional to $V_{ub} \, V_{us}^*$ have been neglected.\\
$O_1$ and $O_2$ are current-current operators,
\begin{subequations}
\label{current_current}
\begin{align}
\label{O1}
O_1 & = ( \bar{c}_\alpha \, \gamma_\mu \, P_L \, b_\beta ) \, ( \bar{s}_\beta \, \gamma^\mu \, P_L \, c_\alpha ) \;, \\
\label{O2}
O_2 & = ( \bar{c} \, \gamma_\mu \, P_L \, b ) \, ( \bar{s} \, \gamma^\mu \, P_L \, c ) \;, 
\end{align}
\end{subequations}
$O_i$ $(i=3,\dots, 6)$ are QCD penguins,
\begin{subequations}
\label{qcd_penguins}
\begin{align}
\label{O34}
O_3 & = ( \bar{s} \, \gamma^\mu \, P_L \, b ) \, \sum_q \, ( \bar{q} \, \gamma^\mu \, P_L \, q ) \;,
\hspace{1cm}
O_4 = ( \bar{s}_\alpha \, \gamma^\mu \, P_L \, b_\beta ) \, \sum_q \, ( \bar{q}_\beta \, \gamma^\mu \, P_L \, q_\alpha ) \;, \\
\label{O56}
O_5 & = ( \bar{s} \, \gamma^\mu \, P_L \, b ) \, \sum_q \, ( \bar{q} \, \gamma^\mu \, P_R \, q) \;,
\hspace{1cm} 
O_6 = ( \bar{s}_\alpha \, \gamma^\mu \, P_L \, b_\beta ) \, \sum_q \, ( \bar{q}_\beta \, \gamma^\mu \, P_R \, q_\alpha ) \;.
\end{align}
\end{subequations}
In \eqref{current_current} - \eqref{qcd_penguins} $P_{R,L} = \displaystyle\frac{1 \pm \gamma_5}{2}$ denote the helicity projectors,  $\alpha$ and $\beta$ are colour indices.
The sum in \eqref{qcd_penguins} runs over the flavours $q = \{ u,d,s,c,b \}$.
The magnetic penguin operators are
\begin{align}
\label{O7}
O_7 & = \frac{e}{16 \, \pi^2} \, \big[ \bar{s} \,\sigma^{\mu\nu} \, ( m_s \, P_L + m_b \, P_R )\,b\big] \, F_{\mu \nu} \;, \\
\label{O8}
O_8 & = \frac{g_s}{16 \, \pi^2} \, \Big[ \bar{s}_{\alpha} \, \sigma^{\mu\nu} \, \Big( \frac{\lambda^a}{2} \Big)_{\alpha \beta} \, (m_s \, P_L + m_b \, P_R) \, b_\beta \Big] \, G^a_{\mu \nu} \;,
\end{align}
and the semileptonic electroweak penguin operators
\begin{align}
\label{O9}
O_9 & = \frac{e^2}{16 \, \pi^2} \, ( \bar{s} \, \gamma^\mu \, P_L \, b) \, ( \bar{\ell} \, \gamma_\mu \, \ell ) \;, \\
\label{O10}
O_{10} & = \frac{e^2}{16 \, \pi^2} \, ( \bar{s} \, \gamma^\mu \, P_L \, b) \, ( \bar{\ell} \, \gamma_\mu \, \gamma_5 \, \ell ) \;.
\end{align}
In previous eqs. $\lambda^a$ are Gell-Mann matrices, $F_{\mu \nu}$ and $G^a_{\mu \nu}$ the electromagnetic and the gluonic field strength tensors, respectively, $e$ and $g_s$ the electromagnetic and strong coupling constants. $m_{b(s)}$ is the $b(s)$ quark mass.\\
The most important operators for $b \to s \, \ell^+ \, \ell^-$ modes are $O_9$ and $O_{10}$.
In general new physics (NP) scenarios other operators can be present, such as those with opposite chirality or those with a different Dirac structure, scalar, pseudoscalar or tensor operators.
In 331 models the tree-level $Z^\prime$ exchange leads to a  modification of the values of the Wilson coefficients $C_{9,10}$:
\begin{align}
\label{C9NP}
C_9^\text{NP} & = C_9^\text{SM} + C_9^{331} \;, \\
\label{C10NP}
C_{10}^\text{NP} & = C_{10}^\text{SM} + C_{10}^{331} \;,
\end{align}
where \cite{Buras:2013dea}
\begin{align}
\label{C9}
\sin^2 \theta_W \, C^{331}_9 & = - \frac{1}{g_\text{SM}^2 \, M_{Z^\prime}^2} \, \frac{\Delta_L^{sb}(Z') \, \Delta_V^{\mu\bar{\mu}}(Z')}{V_{tb} \, V_{ts}^*} \;, \\
\label{C10}
\sin^2 \theta_W \, C^{331}_{10} & = - \frac{1}{g_\text{SM}^2 \, M_{Z^\prime}^2} \, \frac{\Delta_L^{sb}(Z') \, \Delta_A^{\mu\bar{\mu}}(Z')}{V_{tb} \, V_{ts}^*} \;.
\end{align}
In previous eqs. the couplings to leptons are defined as
\begin{align}
\Delta_{V,A}^{\mu\bar{\mu}}(Z') = \Delta_R^{\mu\bar{\mu}}(Z') \pm \Delta_L^{\mu\bar{\mu}}(Z') \;,
\end{align}
where $\Delta_L^{sb}(Z')$ and $\Delta_{R,L}^{\mu\bar{\mu}}(Z')$ are given in eqs. \eqref{C_L_331_interi} and \eqref{Delta_for_leptons}.

The real (imaginary) parts of $C_9^{331}$ and $C_{10}^{331}$ separately depend on the parameters of the model but the ratio $C_9^{331} / C_{10}^{331}$, it is independent of the parameters $\tilde{s}_{13}$, $\tilde{s}_{23}$, $\delta_1$, $\delta_2$ and $M_{Z'}$ \cite{Buras:2023ldz}.
In fact, it only depends on the model variant, and if no $Z - Z'$ mixing is considered, its expression reads: 
\begin{align}
\label{ratio_c9_c10}
\frac{C_9^{331}}{C_{10}^{331}} = \frac{\text{Re}(C_9^{331})}{\text{Re}(C_{10}^{331})} = \frac{\text{Im}(C_9^{331})}{\text{Im}(C_{10}^{331})} = - \frac{\big( \sqrt{3} + 9 \, \beta \big) \, \sin^2 \theta_W - \sqrt{3}}{\big( \sqrt{3} - 3 \, \beta \big) \, \sin^2 \theta_W - \sqrt{3}} \;.
\end{align}
For the four variant in the analysis, without considering $Z - Z'$ mixing, the ratio in \eqref{ratio_c9_c10} has the numerical values:
\begin{align}
\frac{C_9^{331}}{C_{10}^{331}} \hspace{0.5cm} = \hspace{0.5cm}
\begin{cases}
\hspace{0.5cm} - 8.874 \hspace{1cm} (M_1) \\
\hspace{0.5cm} - 2.984 \hspace{1cm} (M_3) \\
\hspace{0.5cm} - 0.004 \hspace{1cm} (M_5) \\
\hspace{0.5cm} + 0.595 \hspace{1cm} (M_7)
\end{cases}
\;.
\end{align}

In the case of  the transition $b \to s \, \nu \, \bar{\nu}$,  the SM effective Hamiltonian  has a simpler structure deriving from penguin and box diagrams \cite{Buras:2020xsm}.
It consists of a single operator
\begin{align}
\label{effective_hamiltonian_b_to_s_nu_nubar}
\mathcal{H}_{b \to s \, \nu \, \bar{\nu}}^\text{SM, eff} = C_L^\text{SM}(b,s) \, O_L + \text{h.c.} \;,
\end{align}
with
\begin{align}
\label{O_L}
O_L & = ( \bar{s} \, \gamma^\mu \, P_L \, b ) \, ( \bar{\nu} \, \gamma_\mu \, P_L \, \nu ) \;.
\end{align}
The SM coefficient $C_L^\text{SM}(b,s)$ depends on the quarks in the initial and in the final state:
\begin{align}
\label{C_L_SM_interi}
C_L^\text{SM}(b,s) & = g_\text{SM}^2 \, \sum_{U = u, c, t} \, V_{Ub} \, V_{Us}^* \, X(x_U) \;,
\end{align}
with $g_\text{SM}^2$ defined in \eqref{gSM2} and $X(x_q)$ the Inami-Lim function depending on the ratio $x_q = m_q^2 / M_W^2$  \cite{Buras:1998raa}.
The dominant contribution comes from the virtual contribution of the top quark and produces
\begin{align}
\label{C_L_SM_interi_max_contribution}
\big| C_L^\text{SM}(b,s) \big| & \simeq \mathcal{O} ( 10^{-8} )  \;.
\end{align}
In this case, the impact of NP could be a modification of the value of $C_L^\text{SM}(b,s)$ or the presence of another operator with opposite chirality of the quark current.
In 331 models one has
\begin{align}
X(x_t) \to X(B_q) = X(x_t) +  \frac{1}{g_\text{SM}^2 \, M_{Z'}^2} \, \frac{\Delta_L^{sb}(Z') \, \Delta_L^{\nu\bar{\nu}}(Z')}{V_{tb} \, V_{ts}^*} \;,
\end{align}
and the NP coefficient becomes
\begin{align}
C_L^\text{NP}(b,s) = C_L^\text{SM}(b,s) + C_L^{331}(b,s) \;,
\end{align}
where \cite{Buras:2012dp}
\begin{align}
C_L^{331}(b,s) = \frac{\Delta_L^{sb}(Z') \, \Delta_L^{\nu\bar{\nu}}(Z')}{M_{Z'}^2} \;.
\end{align}
In the next Section we consider selected FCNC observables that depend on these coefficients within 331 models.

\section{331 models facing new data on selected observables in $b \to s$ processes}
\label{numerics}

\subsection{$b \to s \, \ell^+ \, \ell^-$}

\begin{figure}[!tb]
\begin{center}
\includegraphics[scale=0.45]{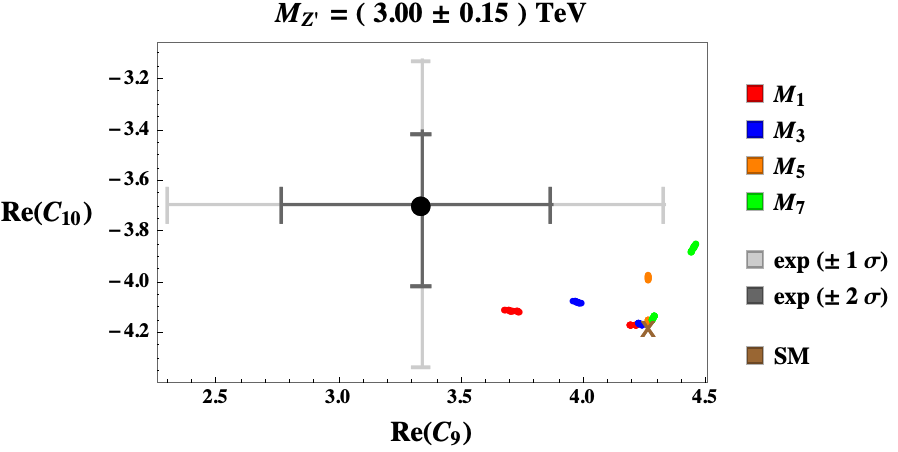}
\hspace{0.4cm}
\includegraphics[scale=0.45]{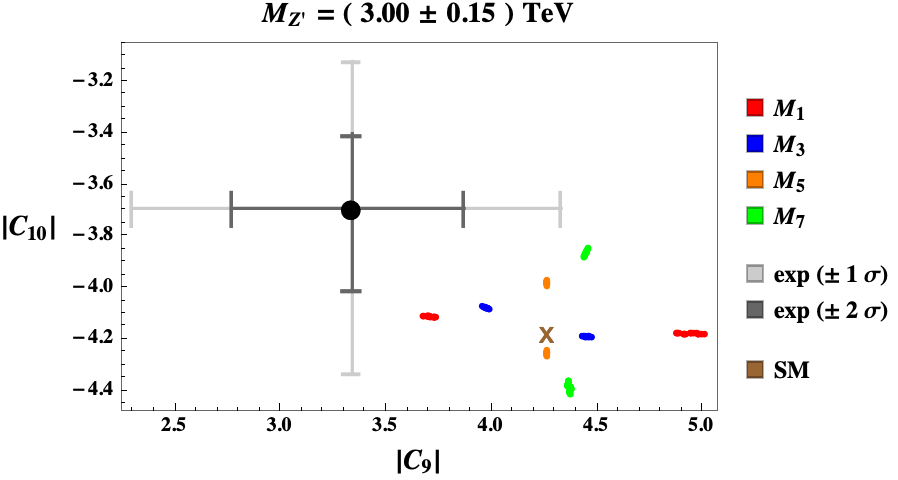}\\
\vspace{0.2cm}
\includegraphics[scale=0.45]{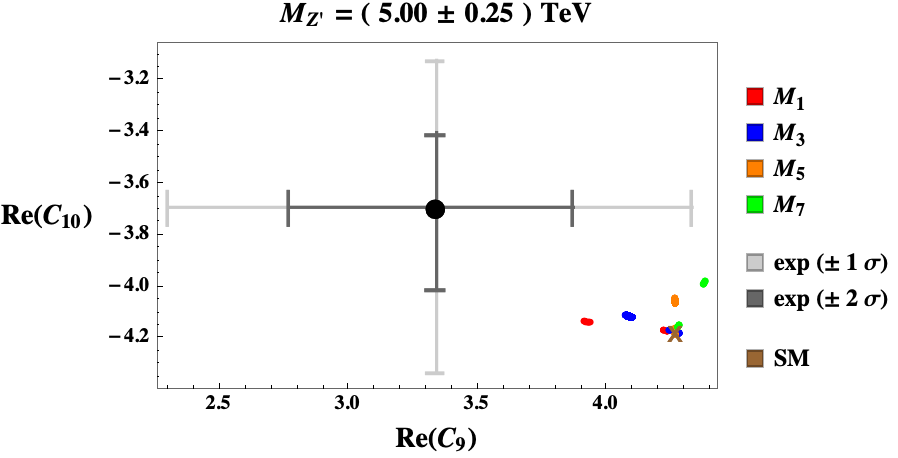}
\hspace{0.4cm}
\includegraphics[scale=0.45]{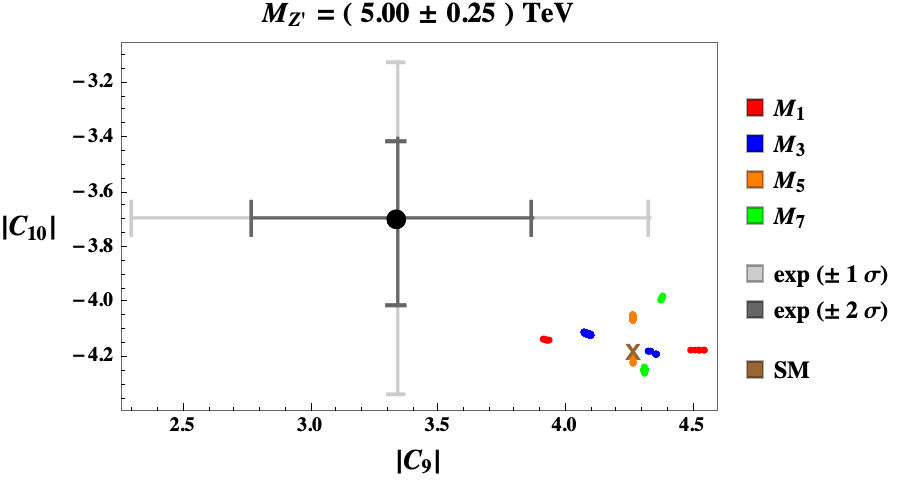}\\
\vspace{0.2cm}
\includegraphics[scale=0.45]{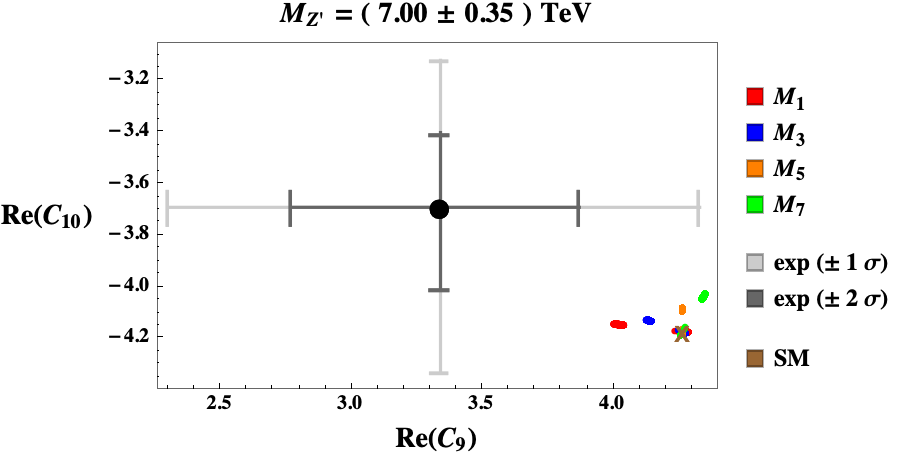}
\hspace{0.4cm}
\includegraphics[scale=0.45]{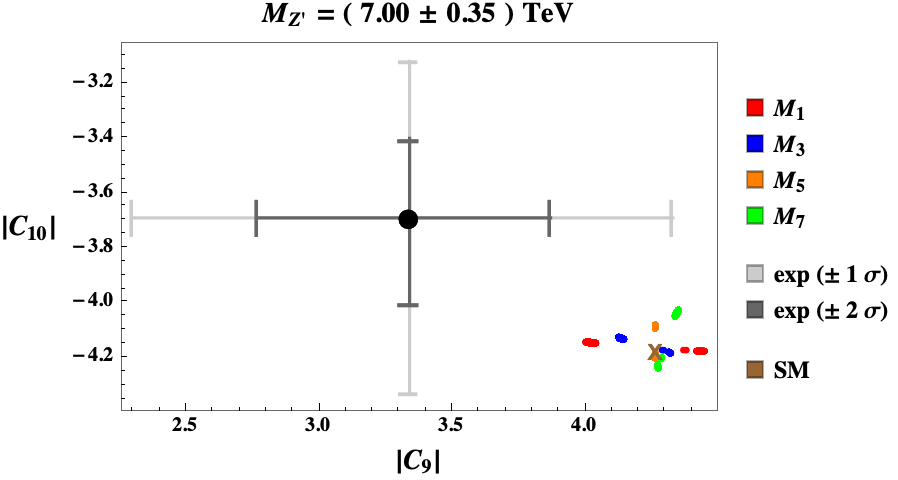}
\end{center}
\caption{
\small
Correlation plot between Wilson coefficients $C_9$ and $C_{10}$ for three selected ranges of $M_{Z'}$.
The plots on the left are obtained assuming that $C_{9,10}$ are real, as assumed in \cite{LHCb:2023gel}, while those on the right take into account the possibility that they have an imaginary part.
The gray bars correspond to the experimental results in \cite{LHCb:2023gel} taken at $1 \, \sigma$ and $2 \, \sigma$ level, while the brown cross represents the SM prediction. 
The other points corresponding  to different 331 variants (different values of the parameter $\beta$) are distinguished by the colours: red ($M_1$), blue ($M_3$), orange ($M_5$) and green ($M_7$).}
\label{c9_vs_c10_diff_MZp}
\end{figure}

\begin{figure}[!tb]
\begin{center}
\includegraphics[scale=0.4]{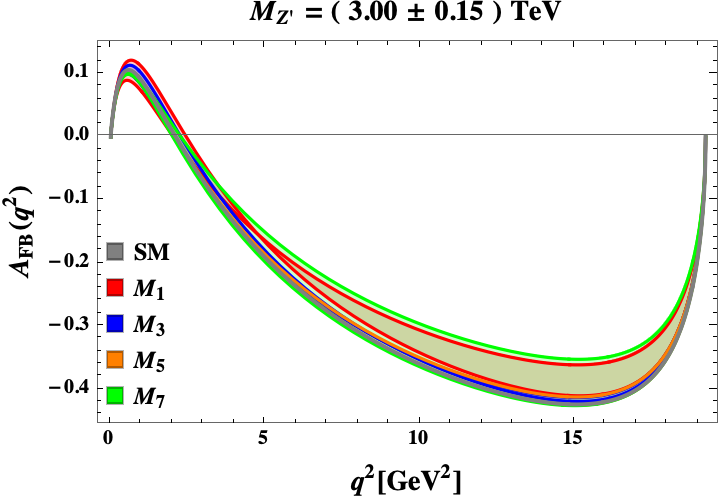}
\hspace{0.3cm}
\includegraphics[scale=0.4]{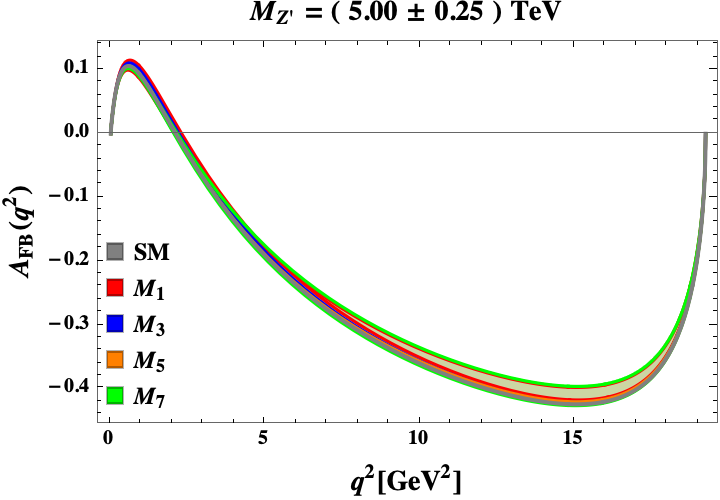}
\hspace{0.3cm}
\includegraphics[scale=0.4]{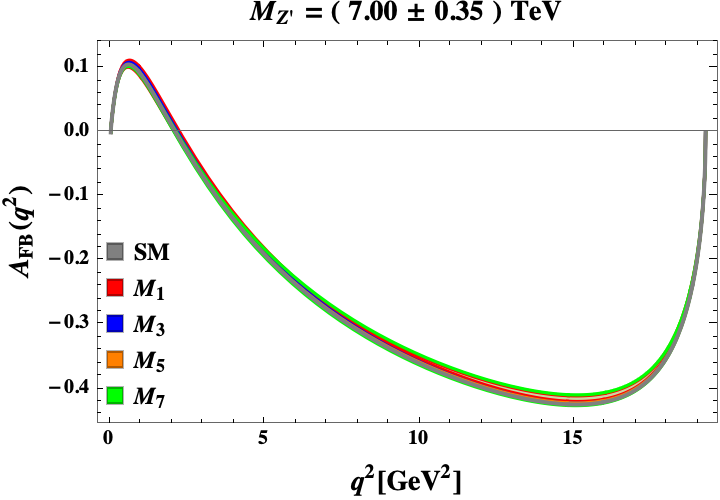}
\end{center}
\vspace{-0.3cm}
\caption{
\small
Forward-backward asymmetry \eqref{A_FB}.
The SM prediction (gray curve) is compared to the four considered 331 models and $M_{Z'} \simeq 3 \, \text{TeV}$ (left), $M_{Z'} \simeq 5 \, \text{TeV}$ (middle) and $M_{Z'} \simeq 7 \, \text{TeV}$ (right panel).
For $M_i$ models, the legend is the same as in Fig. \ref{c9_vs_c10_diff_MZp}.
}
\label{AFB_diff_MZp}
\end{figure}

\begin{figure}[!tb]
\begin{center}
\includegraphics[scale=0.5]{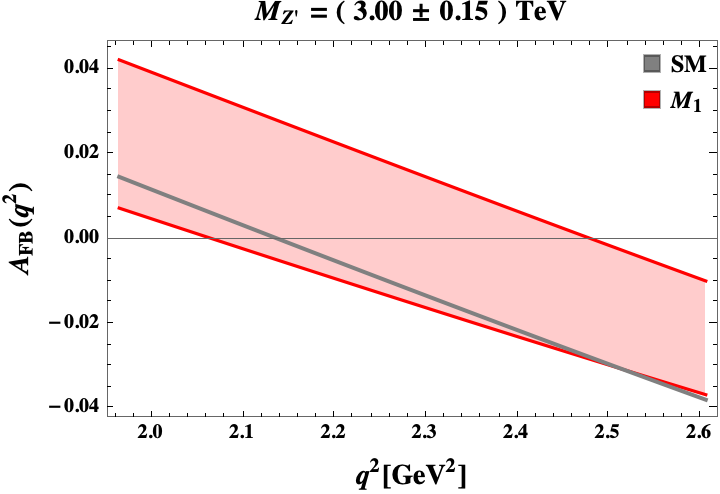}
\hspace{0.2cm}
\includegraphics[scale=0.5]{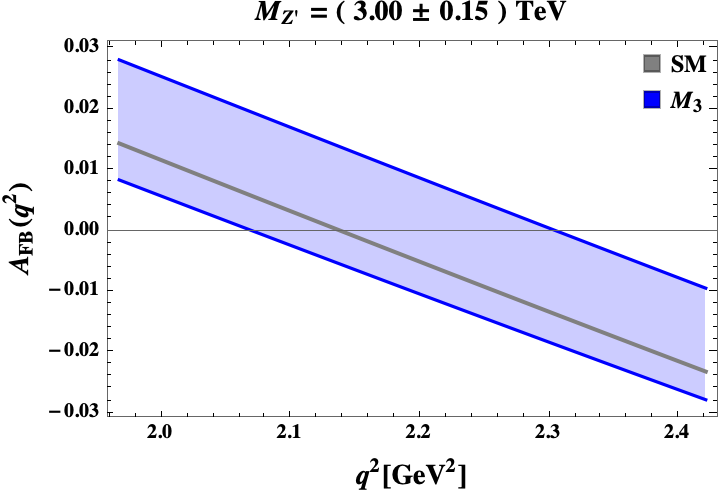}\\
\vspace{0.2cm}
\includegraphics[scale=0.5]{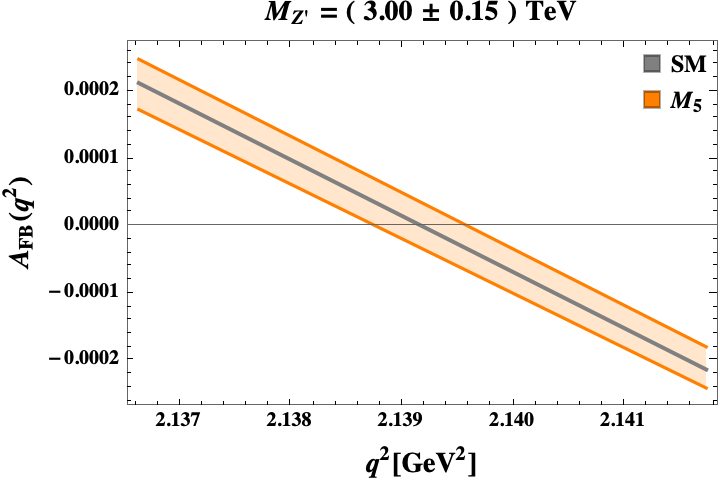}
\hspace{0.2cm}
\includegraphics[scale=0.5]{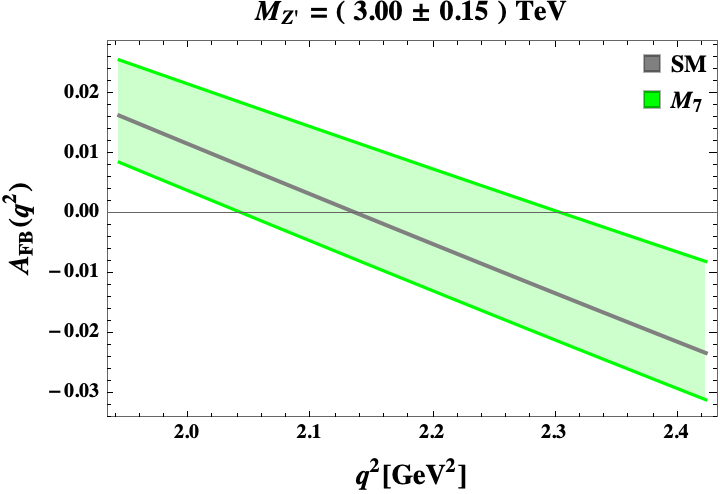}
\end{center}
\vspace{-0.3cm}
\caption{
\small
Zoom of the region close to the zero of the forward-backward asymmetry \eqref{A_FB} for $B \to K^* \, \mu^+ \, \mu^-$.
The SM prediction is compared to $M_1$ (top-left), $M_3$ (top-right), $M_5$ (bottom-left) and $M_7$ (bottom-right) in the case $M_{Z'} \simeq 3 \, \text{TeV}$.
Same legend as in Fig. \ref{AFB_diff_MZp}.
}
\label{AFB_diff_MZp_3TeV}
\end{figure}

\begin{figure}[!tb]
\begin{center}
\includegraphics[scale=0.5]{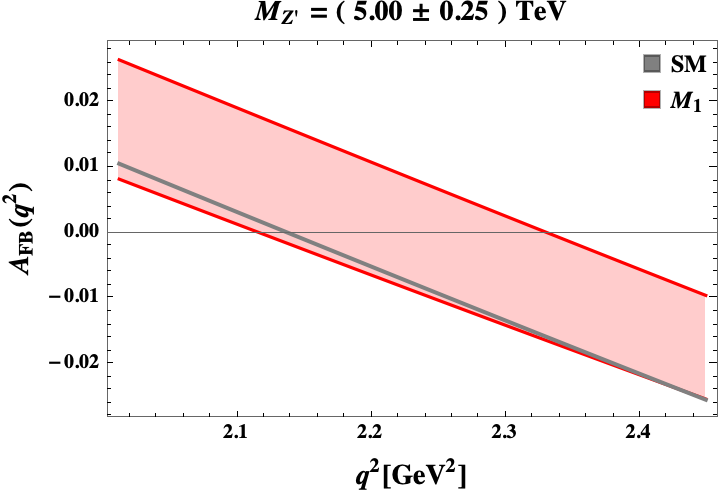}
\hspace{0.2cm}
\includegraphics[scale=0.5]{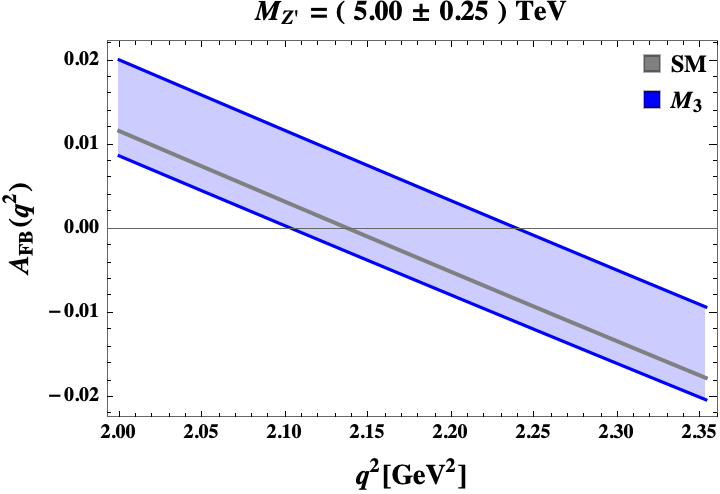}\\
\vspace{0.2cm}
\includegraphics[scale=0.5]{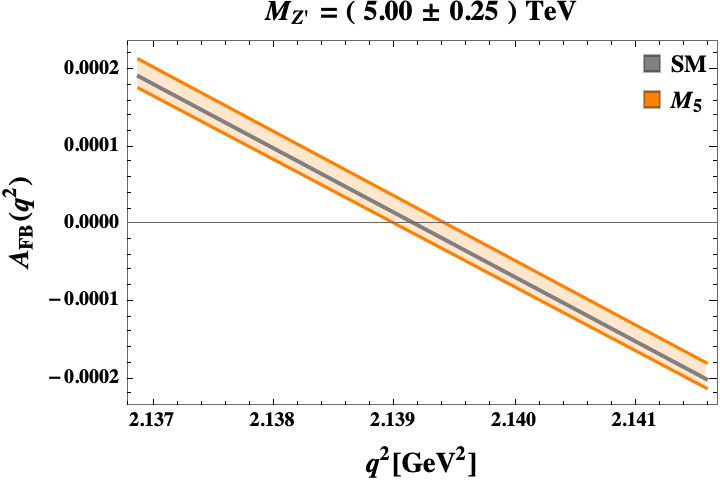}
\hspace{0.2cm}
\includegraphics[scale=0.5]{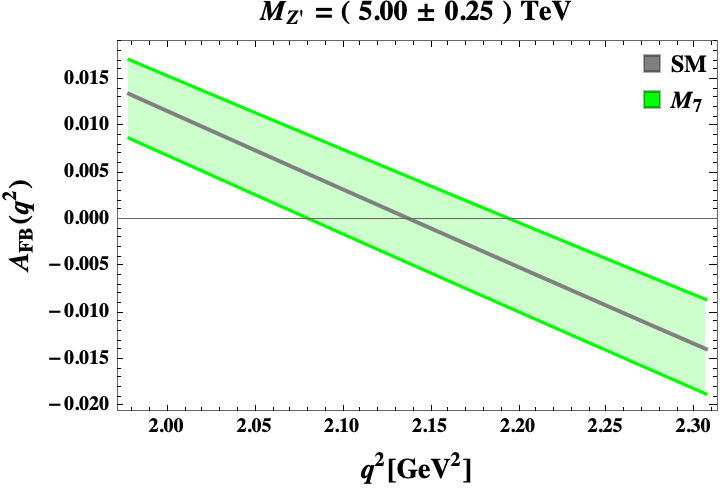}
\end{center}
\vspace{-0.3cm}
\caption{
\small
Zoom of the region close to the zero of the forward-backward asymmetry \eqref{A_FB} for $B \to K^* \, \mu^+ \, \mu^-$.
The SM prediction is compared to $M_1$ (top-left), $M_3$ (top-right), $M_5$ (bottom-left) and $M_7$ (bottom-right) in the case $M_{Z'} \simeq 5 \, \text{TeV}$.
Same legend as in Fig. \ref{AFB_diff_MZp}.
}
\label{AFB_diff_MZp_5TeV}
\end{figure}

\begin{figure}[!tb]
\begin{center}
\includegraphics[scale=0.5]{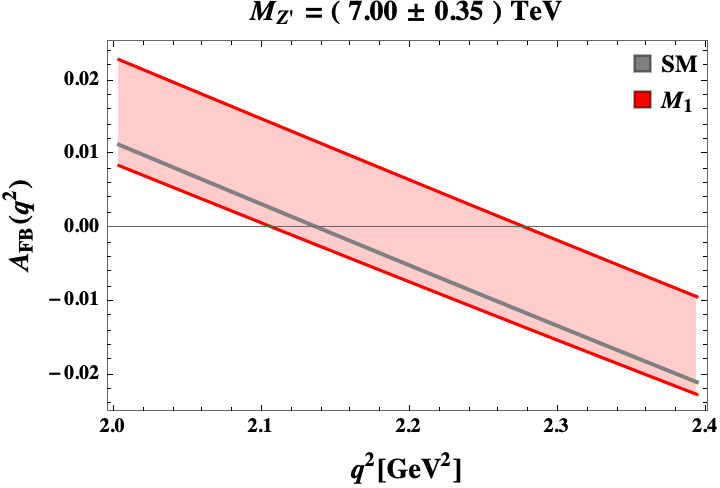}
\hspace{0.2cm}
\includegraphics[scale=0.5]{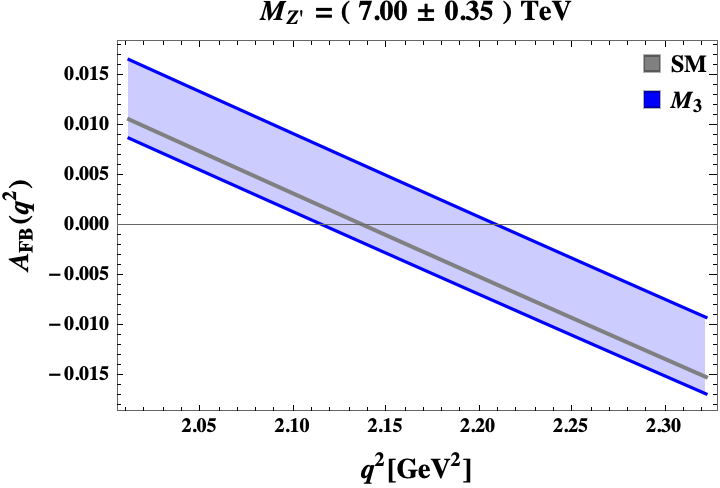}\\
\vspace{0.2cm}
\includegraphics[scale=0.5]{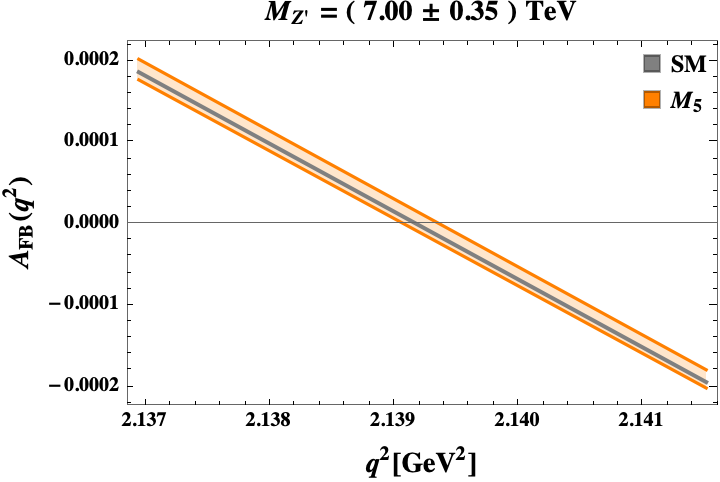}
\hspace{0.2cm}
\includegraphics[scale=0.5]{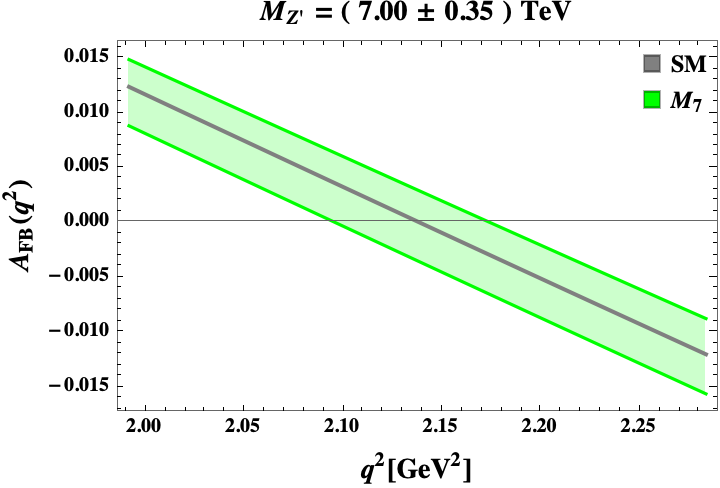}
\end{center}
\vspace{-0.3cm}
\caption{
\small
Zoom of the region close to the zero of the forward-backward asymmetry \eqref{A_FB} for $B \to K^* \, \mu^+ \, \mu^-$.
The SM prediction is compared to $M_1$ (top-left), $M_3$ (top-right), $M_5$ (bottom-left) and $M_7$ (bottom-right) in the case $M_{Z'} \simeq 7 \, \text{TeV}$.
Same legend as in Fig. \ref{AFB_diff_MZp}.
}
\label{AFB_diff_MZp_7TeV}
\end{figure}

A recent study of the LHCb Collaboration \cite{LHCb:2023gel} has provided the values of the Wilson coefficients $C_9^{(\prime)}$ and $C_{10}^{(\prime)}$ fitted from the amplitude analysis of the mode $B \to K^{0*} \, \mu^+ \, \mu^-$.
It is interesting to consider how 331 models face these new experimental data. 
According to \cite{LHCb:2023gel}, no sensitivity to the imaginary part of the Wilson coefficients can be achieved treating simultaneously $B^0$ and $\bar{B}^0$ decays, so that all coefficients are assumed to be real.
Since  in 331 models the coefficients do have an imaginary part, we compare the LHCb findings both to the real part of $C_{9,\,10}$, both to their moduli.
In this way we can also appreciate the role of the imaginary part. 
The results are displayed in Fig. \ref{c9_vs_c10_diff_MZp}, comparing the real parts of the coefficients to data in the plots in the left column and their moduli in plots in the right column.
Both the SM and 331 models can reproduce the data at $2 \, \sigma$. 
The variant that performs better on the basis of these observables alone is $M_1$.
For $B \to K^* \, \mu^+ \, \mu^-$ another observable can be considered, the  forward-backward asymmetry $A_\text{FB}(q^2)$
\begin{align}
\label{A_FB}
A_\text{FB}(q^2) = \bigg[ \int_0^1 \, \text{d} \cos \theta \, \frac{\text{d}^2 \Gamma}{\text{d} q^2 \, \text{d} \cos \theta} - \int_{-1}^0 \, \text{d} \cos \theta \, \frac{\text{d}^2 \Gamma}{\text{d} q^2 \, \text{d} \cos \theta} \bigg] \bigg/ \frac{\text{d} \Gamma}{\text{d} q^2} \;,
\end{align}
where $\theta$ is the angle between the positive charged muon and the $B$ meson in the $\mu^+ \, \mu^-$ pair rest frame, and $q^2$ is the invariant mass of the muon pair \cite{Altmannshofer:2008dz,Beneke:2000wa}.
In SM this observable has a zero, a value $q^2 = s_0$ such that $A_{FB}(s_0) = 0$.
The value of $s_0$ is almost independent of the model chosen for the form factors \cite{Beneke:2000wa}, so it represents a clean observable to probe possible NP contribution.
We consider the model proposed in \cite{Ball:2004rg} neglecting the form factors uncertainties.
There are NP scenarios that predict either that there is no zero, or that it is displaced with respect to SM.
$s_0 $ is defined by the relation:
\begin{align}
\label{zero_AFB}
& \big| C_{7} \big| \, \cos \text{Arg}(C_{10}) \, m_b \, \big\{ ( M_{B} - M_{K^*} ) \, V(s_0) \, T_2(s_0) + 2 \, ( M_{B} + M_{K^*} ) \, A_1(s_0) \, T_1(s_0) \big\} + \notag \\[0.5ex]
& \hspace{1cm} + \big| C_{9} \big| \, \cos \big[ \text{Arg}(C_{10}) - \text{Arg}(C_{9}) \big] \, 2 \, s_0 \, V(s_0) \, A_1(s_0) = 0 \;,
\end{align}
The functions $V$, $A_1$, $T_1$ and $T_2$ in \eqref{zero_AFB} are form factors that parametrize the $B \to K^*$ matrix element of the operators in the effective Hamiltonian.
Their definition can be found in Appendix \ref{app_A}.
Indeed, since we are interested in the position of the zero that has a reduced dependence on the form factors, this assumption makes more transparent the comparison with 331 models where the location of the zero is more uncertain due to the variation of the parameters within their allowed ranges.
Fig. \ref{AFB_diff_MZp} shows the results for this observable in full range of $q^2$, comparing the SM prediction (depicted in green) to the four considered 331 variants.
The range of $q^2$ close to the zero is enlarged in Figs. \ref{AFB_diff_MZp_3TeV}, \ref{AFB_diff_MZp_5TeV} and \ref{AFB_diff_MZp_7TeV}.
It can be observed that, except for $M_5$, in all variants a shift with respect to SM is possible.
The largest deviations are found in $M_1$.
The results are in Tab. \ref{tabellazeri}.
\begin{table}[!tb]
\centering
\small
\begin{tabular}{|c|c|c|c|}
\hline
$M_{Z'}$ (TeV) & \multicolumn{1}{c|}{$3.00 \pm 0.15$} & \multicolumn{1}{c|}{$5.00 \pm 0.25$} & \multicolumn{1}{c|}{$7.00 \pm 0.35$} \\
\hline
$M_1$ & $s_0 \in [2.0677,2.4821] \, \text{GeV}^2$ & $s_0 \in [2.1181,2.3325] \, \text{GeV}^2$ & $s_0 \in [2.1089,2.2798] \, \text{GeV}^2$ \\
\hline
$M_3$ & $s_0 \in [2.0708,2.3059] \, \text{GeV}^2$ & $s_0 \in [2.1051,2.2410] \, \text{GeV}^2$ & $s_0 \in [2.1174,2.2112] \, \text{GeV}^2$ \\
\hline
$M_5$ & $s_0 \in [2.1388,2.1396] \, \text{GeV}^2$ & $s_0 \in [2.1390,2.1394] \, \text{GeV}^2$ & $s_0 \in [2.1391,2.1394] \, \text{GeV}^2$ \\
\hline
$M_7$ & $s_0 \in [2.0447,2.3075] \, \text{GeV}^2$ & $s_0 \in [2.0821,2.1965] \, \text{GeV}^2$ & $s_0 \in [2.0965,2.1748] \, \text{GeV}^2$ \\
\hline
\end{tabular}
\vspace{0.2cm}
\caption{
\small
Ranges of the zero $s_0$ of the  forward-backward asymmetry in $B \to K^* \, \mu^+ \, \mu^-$ in the four variants of the 331 models considered in this paper.
}
\label{tabellazeri}
\end{table}

\subsection{$b \to s \, \nu \, \bar{\nu}$}

The processes $B \to K^{(*)} \, \nu \, \bar{\nu}$ are theoretically clean.
In the SM, their branching ratios  are predicted \cite{Becirevic:2023aov}
\begin{align}
\label{KSM}
\mathcal{B}(B^+ \to K^+ \, \nu \, \bar{\nu})_\text{SM} & = ( 5.22 \pm 0.15 \pm 0.28 ) \times 10^{-6} \;, \\
\label{KstarSM}
\mathcal{B}(B^+ \to K^{*+} \, \nu \, \bar{\nu})_\text{SM} & = ( 11.27 \pm 1.38 \pm 0.62 ) \times 10^{-6} \;,
\end{align}
updating previous results \cite{Colangelo:1996ay,Buras:2014fpa,Altmannshofer:2009ma,Blake:2016olu,Parrott:2022zte}.
On the other hand, due to the neutrino pair in the final state, they are experimentally changelling.
Recently, the Belle II Collaboration has provided the measurement \cite{Belle-II:2023esi}:
\begin{align}
\label{bellenunu}
\mathcal{B}(B^+ \to K^+ \, \nu \, \bar{\nu})_\text{exp} = ( 2.7 \pm 0.5 \, (\text{stat} ) \pm 0.5 \, (\text{syst}) ) \times 10^{-5}
\end{align}
displaying a tension with the SM result \eqref{KSM}.
This tension has already triggered a number of analyses \cite{Bause:2023mfe,Allwicher:2023xba,Berezhnoy:2023rxx,Datta:2023iln,McKeen:2023uzo,Fridell:2023ssf}.

\begin{figure}[!tb]
\begin{center}
\includegraphics[scale=0.4]{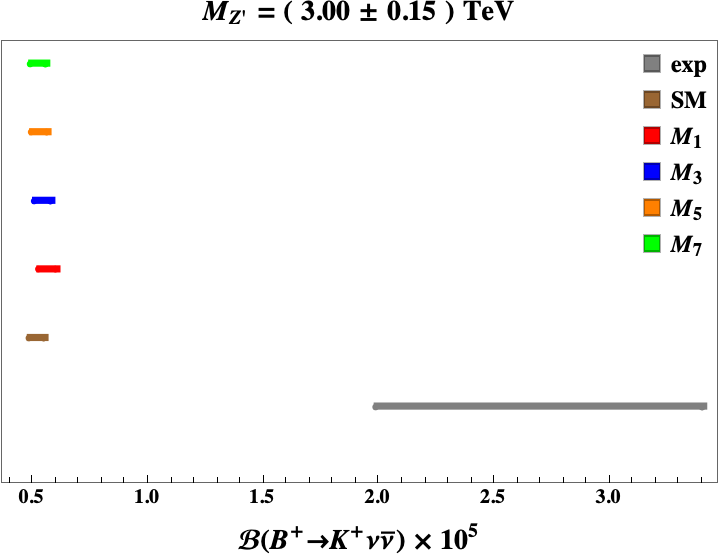}
\hspace{0.3cm}
\includegraphics[scale=0.4]{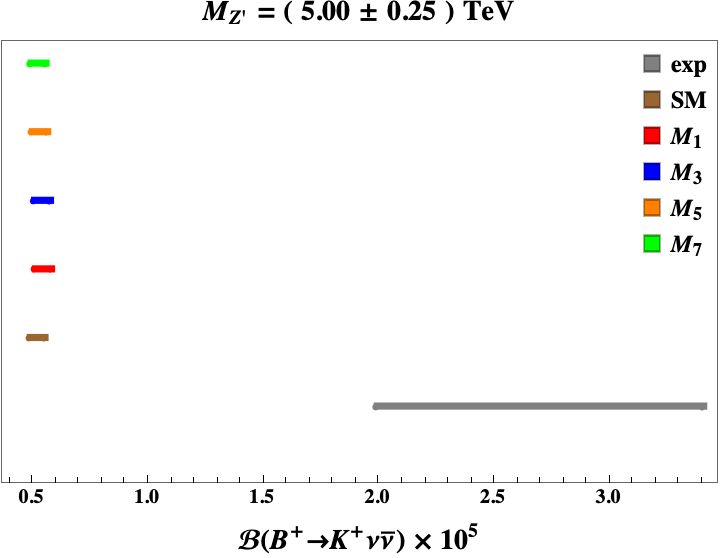}
\hspace{0.3cm}
\includegraphics[scale=0.4]{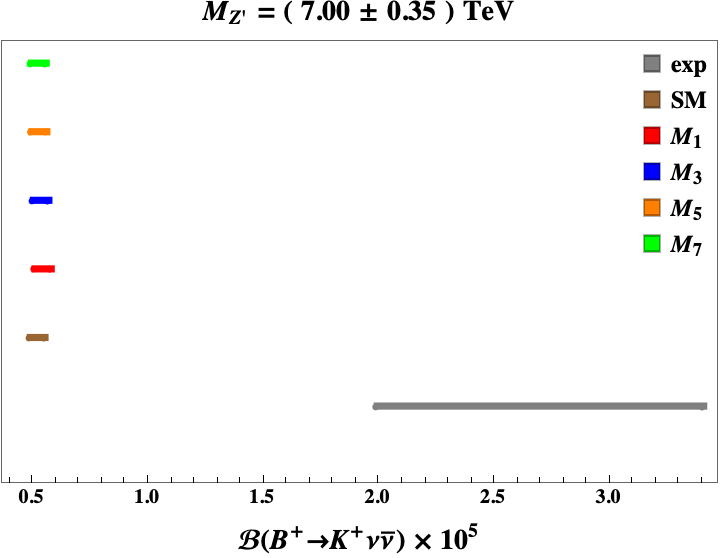}
\end{center}
\vspace{-0.3cm}
\caption{
\small
Branching ratio $\mathcal{B} ( B^+ \to K^+ \, \nu \, \bar{\nu} )$.
Experimental data (blue line) and SM prediction (green line) are compared to the four different 331 models considered in this study.
Same legend as in Fig. \ref{AFB_diff_MZp}.}
\label{br_diff_MZp}
\end{figure}

Considering the 331 models, the branching ratio reads
\begin{align}
\label{br_NP}
\mathcal{B} ( B^+ \to K^+ \, \nu \, \bar{\nu} )_\text{NP} = \left| \frac{C_L^\text{NP}(b,s)}{C_L^\text{SM}(b,s)} \right|^2 \, \mathcal{B} ( B^+ \to K^+ \, \nu \, \bar{\nu} )_\text{SM} \;.
\end{align}
Using the results \eqref{KSM} and \eqref{bellenunu} in eq. \eqref{br_NP}, we find that, to reproduce the data, the NP contribution should enhance $C_L^\text{NP}(b,s)$,
\begin{align}
\left| \frac{C_L^\text{NP}(b,s)}{C_L^\text{SM}(b,s)} \right| \simeq 2.3 \pm 0.3 \;,
\end{align}
a relation not satisfied in 331 models, as argued from Fig. \ref{br_diff_MZp}.

In Tab. \ref{tabella_CL} we compare our results to those obtained in \cite{Buras:2012dp,Colangelo:2021myn}.

%{
%\color{red}
The new results decrease as the mass $M_{Z'}$ increases but not as fast as in the previous analyses.
In addition, the values are always larger than $1$ at $1 \, \sigma$ level.
%}

\begin{table}[!tb]
\centering
\footnotesize
\begin{tabular}{|c|c|c|c|c|c|c|}
\hline
\multicolumn{7}{|c|}{$\big| C_L^\text{NP}(b,s) / C_L^\text{SM}(b,s) \big|^2$} \\
\hline
$M_{Z'}$ (TeV) & \multicolumn{2}{c|}{$3.00 \pm 0.15$} & \multicolumn{2}{c|}{$5.00 \pm 0.25$} & \multicolumn{2}{c|}{$7.00 \pm 0.35$} \\
\cline{2-7}
Variants & \cite{Buras:2012dp,Colangelo:2021myn} & Our result & \cite{Buras:2012dp,Colangelo:2021myn} & Our result & \cite{Buras:2012dp,Colangelo:2021myn} & Our result \\
\hline
$M_1$ & $1.003 \pm 0.105$ & $1.090 \pm 0.024$ & $1.001 \pm 0.063$ & $1.047 \pm 0.020$ & $1.001 \pm 0.046$ & $1.028 \pm 0.022$ \\
\hline
$M_3$ & $1.001 \pm 0.067$ & $1.049 \pm 0.021$ & $1.000 \pm 0.040$ & $1.024 \pm 0.020$ & $1.000 \pm 0.029$ & $1.017 \pm 0.015$ \\
\hline
$M_5$ & $1.000 \pm 0.033$ & $1.022 \pm 0.013$ & $1.000 \pm 0.020$ & $1.011 \pm 0.011$ & $1.000 \pm 0.014$ & $1.008 \pm 0.008$ \\
\hline
$M_7$ & $1.000 \pm 0.021$ & $1.013 \pm 0.009$ & $1.000 \pm 0.013$ & $1.008 \pm 0.006$ & $1.000 \pm 0.009$ & $1.005 \pm 0.005$ \\
\hline
\end{tabular}
\vspace{0.2cm}
\caption{
\small
Enhancement of $C_L^\text{NP}(b,s)$ for processes mediated by $b \to s \, \nu \, \bar{\nu}$ transition in the 331 variants considered in this study with respect to SM prediction.
Comparison to the results in \cite{Buras:2012dp,Colangelo:2021myn} is also provided.
}
\label{tabella_CL}
\end{table}

\section*{Conclusions}

The important role played by FCNC processes in the search for BSM physics is due to their sensitivity to the virtual contributions of heavy particle exchanges.
Exclusive processes such as the purely leptonic $B_s \to \mu^+ \, \mu^-$ and the semileptonic ones $B \to K^{(*)} \, \ell^+ \, \ell^-$, with $\ell=e, \mu$, allow to investigate possible effects of BSM through several observables.

In this analysis we have considered the 331 models, based on a larger gauge group which includes the SM one.
In such models, FCNC processes are mediated by a new heavy neutral gauge boson $Z'$ tree-level exchange.

A new approach has been used to constrain the set of the 331 parameters which enter in the observables relative to $B \to K^* \, \mu^+ \, \mu^-$ and $B \to K^{(*)} \, \nu \, \bar{\nu}$.
Among the possible 24 models, only four have been selected, showing that for the first process 331 models can perform better than SM.
On the other hand, for the modes with neutrinos in the final state, the most recent experimental result are in tension both with SM and 331 models.

\section*{Acknowledgements}

I thank F. De Fazio and P. Colangelo for discussions.
This study has been  carried out within the INFN project (Iniziativa Specifica) SPIF, and it has been partly funded by the European Union – Next Generation EU through the research grant number P2022Z4P4B “SOPHYA - Sustainable Optimised PHYsics Algorithms: fundamental physics to build an advanced society" under the program PRIN 2022 PNRR of the Italian Ministero dell’Università e Ricerca (MUR).

%\newpage

\appendix

\section{Form factor parametrization}
\label{app_A}

For $B(p) \to K^*(p', \epsilon)$ transitions the hadronic matrix elements in the decay amplitudes are parametrized in terms of seven form factors:
\begin{subequations}
\begin{align}
\braket{K^*(p',\epsilon) | \bar{s} \, \gamma^\mu \, b | B(p) } & = - \frac{2 \, V(q^2)}{M_{B} + M_{K^*}} \, i \, \varepsilon^{\mu\nu\alpha\beta} \, \epsilon_\nu^* \, p_\alpha \, p'_\beta \;, \\[1ex]
\braket{K^*(p',\epsilon) | \bar{s} \, \gamma^\mu \, \gamma_5 \, b | B(p) } & = 2 \, M_{K^*} \, \frac{\epsilon^* \cdot q}{q^2} \, q^\mu \, A_0(q^2) + ( M_{B} + M_{K^*} ) \, \bigg( \epsilon^{*\mu} - \frac{\epsilon^* \cdot q}{q^2} \, q^\mu \bigg) \, A_1(q^2) + \notag \\[1ex]
& \hspace{1cm} - \frac{\epsilon^* \cdot q}{M_{B} + M_{K^*}} \, \bigg( p^\mu + p^{\prime\mu} - \frac{M_{B}^2 - M_{K^*}^2}{q^2} \, q^\mu \bigg) \, A_2(q^2) \;, \\[1ex]
\braket{K^*(p',\epsilon) | \bar{s} \, \sigma^{\mu\nu} \, b | B(p) } & = \frac{\epsilon^* \cdot q}{( M_{B} + M_{K^*} )^2} \, \varepsilon^{\mu\nu\alpha\beta} \, p_\alpha \, p'_\beta \, T_0(q^2) + \notag \\[1ex]
& \hspace{1cm} + \varepsilon^{\mu\nu\alpha\beta} \, p_\alpha \, \epsilon^*_\beta \, T_1(q^2) + \varepsilon^{\mu\nu\alpha\beta} \, p'_\alpha \, \epsilon^*_\beta \, T_2(q^2) \;, \\[1ex]
\braket{K^*(p',\epsilon) | \bar{s} \, \sigma^{\mu\nu} \, \gamma_5 \, b | B(p) } & = i \, \frac{\epsilon^* \cdot q}{( M_{B} + M_{K^*} )^2} \, ( p_\mu \, p'_\nu - p_\nu \, p'_\mu ) \, T_0(q^2) + \notag \\[1ex]
& \hspace{1cm} + i \, ( p_\mu \, \epsilon^*_\nu - p_\nu \, \epsilon^*_\mu ) \, T_1(q^2) + i \, ( p'_\mu \, \epsilon^*_\nu - p'_\nu \, \epsilon^*_\mu ) \, T_2(q^2) \;,
\end{align}
\end{subequations}
where $q = p - p'$, $\epsilon$ is the polarization of $K^*$ and it is satisfied the condition
\begin{align}
A_0(0) = \frac{M_{B} + M_{K^*}}{2 \, M_{K^*}} \, A_1(0) - \frac{M_{B} - M_{K^*}}{2 \, M_{K^*}} \, A_2(0) \;.
\end{align}

\nocite{}

\bibliographystyle{JHEP}
\bibliography{biblio_particles}

\end{document}